\documentclass[preprint]{aastex}
\usepackage{subfigure}
\begin{document}

\title{ Dissection of H$\alpha$ Emitters : Low-z Analogs of $z>4$ Star-Forming Galaxies }

\author{ Hyunjin Shim\altaffilmark{1, 2} \&
         Ranga-Ram Chary\altaffilmark{3} 
}
\altaffiltext{1}{Department of Earth Science Education, 
Kyungpook National University, Korea; hjshim@knu.ac.kr}
\altaffiltext{2}{Spitzer Science Center,
California Institute of Technology, MS 220-6, Pasadena, CA 91125 }
\altaffiltext{3}{U.S. Planck Data Center,
California Institute of Technology, MS 220-6, Pasadena, CA 91125 }

\begin{abstract}
Strong H$\alpha$ Emitters (HAEs) dominate the $z\sim4$ 
Lyman-break galaxy population. We have identified local analogs of these HAEs 
using the Sloan Digital Sky Survey (SDSS). 
  At $z<0.4$, only 0.04\,\% of galaxies are classified as HAEs 
 with H$\alpha$ equivalent widths ($\gtrsim500\,\mbox{\AA}$) comparable
 to that of $z\sim4$ HAEs. Local HAEs have lower stellar mass
 and lower ultraviolet (UV) luminosity than $z\sim4$ HAEs, yet 
 the H$\alpha$-to-UV luminosity ratio as well as their specific star-formation rate
is consistent with that of $z\sim4$ HAEs indicating that they are scaled down versions of 
high-z star-forming galaxies. 
  Compared to the previously studied local analogs of $z\sim2$ Lyman break galaxies
 selected using rest-frame UV, local HAEs show 
 similar UV luminosity surface density, weaker D$_{n}(4000)$ breaks, 
lower metallicity and lower stellar mass.
 This supports the idea that local HAEs are less evolved galaxies than the traditional Lyman break analogs. 
 We are not able to constrain if the star-formation history in
 local HAEs is powered by mergers or by cosmological cold flow accretion.
  However, in the stacked spectrum, local HAEs show a strong HeII\,$\lambda4686$ emission line 
suggesting a population of young ($<$10\,Myr), hot, massive stars 
similar to that seen in some Wolf-Rayet galaxies.
 Low [NII]/[OIII] line flux ratios imply that local HAEs are inconsistent with being
 systems that host bright AGN. Instead, it is highly likely that  
 local HAEs are galaxies with an elevated ionization parameter, 
 either due to a high electron density or
 large escape fraction of hydrogen ionizing photons 
 as in the case for Wolf-Rayet galaxies. 
\end{abstract}

\keywords{cosmology: observation -- galaxies: evolution -- galaxies: starburst 
 -- galaxies: low-redshift }

\section{Introduction}

 In attempts to understand the cosmic star formation history over cosmic time,
 most conventional approaches select high-redshift star-forming galaxies
 based on different unique signatures in their spectral energy distribution. 
Well-known examples
are the Lyman-break in the ultraviolet (UV) continuum, Lyman-$\alpha$ emission line, 
the Balmer-break in evolved stellar populations
and more recently, the shape of the observed far-infrared spectral 
energy distribution which shows a peak between $\sim$60 and 100$\mu$m.
 The next step is to investigate their physical properties 
 such as star formation rates, stellar mass and stellar populations
 using yardsticks calibrated in the local universe 
 for which higher signal to noise and higher spatial resolution data
 are available, compared to the high-redshift galaxies considered. 
 By studying the local counterparts of high-redshift galaxies,
it is possible to understand the physical trigger for star-formation
and its temporal evolution, and how physical parameters in the interstellar medium
 affect observables that trace the star formation activity. 

 One well known study
 is the investigation of local UV-luminous galaxies as analogs of 
 high-redshift Lyman-break galaxies.
 Heckman et al.(2005) have studied the properties of FUV-luminous
 galaxies at $z\lesssim0.3$ using the combination of 
Galaxy Evolution Explorer (GALEX) and Sloan Digital Sky Survey (SDSS) data. 
 The FUV luminosities of the selected galaxies are greater than 
$2\times10^{10}\,L_{\odot}$,
 comparable to that of $z\sim2-3$ Lyman break galaxies (LBGs). 
 The number density of such UV-luminous galaxies (UVLGs) is 0.85\,\% of all galaxies
 at $0.0<z<0.3$ (Hoopes et al. 2007). Hoopes et al.(2007) have shown that
 the specific star formation rate of UVLGs is correlated with their
 UV surface brightness, and the subset of compact UVLGs with the highest
 UV surface brightness (i.e., $I_{FUV}>10^{9} L_{\odot}$\,kpc$^{-2}$; supercompact UVLGs) 
 are likely to be the local counterparts to high-redshift LBGs 
 in terms of their UV surface brightness, star formation rate, metallicity, 
 and stellar mass. 

 These supercompact UVLGs, later denoted as Lyman break Analogs (LBAs), 
 have been studied in detail in the radio (Basu-Zych et al. 2007)
 and infrared (Overzier et al. 2009; Overzier et al. 2011) which
 helps provide a window into high-redshift star forming environments in the local Universe.
 For example, high spatial resolution studies of the morphologies of LBAs show clear evidence of 
 merger-induced star formation (Overzier et al. 2008, 2010). 
 In contrast, the signatures of mergers are not visible in the rest-frame UV observation of
 high-redshift galaxies. Ravindranath et al. (2006) demonstrated that 
only $\sim$30\% of $z>3$ galaxies 
show signatures of mergers. It is plausible that this derived fraction is a lower limit due to
surface brightness dimming (Overzier et al. 2010). However, there is a theoretical suggestion
 that high-redshift star formation is
 preferentially governed by a supply of cold gas through cold flow accretion
 rather than merging (Dekel et al. 2009). The mechanism may not be valid
 in the local universe due to the fact that the halo masses are much larger and the gas gets shock
heated before it falls within the virial radius of the halo.
 Thus, the `local counterparts' of high-redshift star-forming galaxies may not necessarily
have a similar source of gas supply as distant galaxies. 
 However, their similarity in observable properties are useful for evaluating if the derived physical quantities
for high redshift galaxies, such as star formation rates and dust obscuration are accurate.
 Thus, the key question is whether the UV-selected LBAs are
 accurate local analogs for high-redshift star-forming galaxies. 

 Recently, we have discovered that
 at least 70\,\% of spectroscopically confirmed Lyman-break
galaxies at $3.8<z<5.0$ show a flux density excess in \textit{Spitzer} 3.6\,$\mu$m 
 over the stellar continuum, which is due to redshifted H$\alpha$ emission
 (Shim et al. 2011, H$\alpha$ Emitters; hereafter HAEs).
 The result is striking for two reasons; one is that strong nebular emission lines
 do affect broad-band photometry and photometry-based SED 
fitting (e.g. Chary et al. 2005, Zackrisson et al. 2008, Capak et al. 2011, 
Schaerer \& de Barros 2009), and two, 
 high-redshift galaxies show strong H$\alpha$ emission which implies a star-formation rate
that is as high, if not higher than that inferred from their large UV luminosities.
 The estimated H$\alpha$ equivalent widths (EWs) are in the range of $140-1700\,\mbox{\AA}$, 
 rarely seen in surveys of local galaxies. 
 Despite the selection bias in favor of emission line sources that is
 associated with spectroscopic confirmation of Lyman-break galaxies, 
 the value of 70\,\% is unusually high. It suggests that there is a physical reason
 for this `strong H$\alpha$ phase' that was previously not known. 
 The high H$\alpha$-to-UV ratio and large H$\alpha$ EW compared to the ages of 
the stellar population therein
suggest that at least 50\,\% of $z\sim4$ HAEs
 prefer an extended star formation timescale rather than a burst-like, short ($\sim$100\,Myr) timescale. 
 Therefore, such extended star formation histories should be included 
 in the selection criteria for local counterparts of high-redshift star-forming galaxies.

 Large H$\alpha$ EW galaxies can be easily identified among local samples.
 In this paper, we selected local galaxies ($z<0.4$) with
 H$\alpha$ EW$>$500\,$\mbox{\AA}$ from the Sloan Digital Sky Survey (SDSS).
 We use the measured properties of these local HAEs, 
 to understand the properties of 
 $z\gtrsim4$ HAEs that dominate the high-redshift star-forming galaxy population.
 We investigate possible reasons for the origin of strong H$\alpha$ emission in 
star-forming galaxies (presence of active galactic nuclei, low metallicity, 
extended star formation histories,
 and dust extinction) by using spectroscopically derived values 
 such as line indices and metallicity, as well as UV-to-IR flux ratios.  
 A comparison between properties of HAEs and LBAs is discussed, 
and the implications of our improved understanding of local HAEs 
 to high-redshift star formation is presented. Throughout this paper, 
we use a cosmology with
 $\Omega_{{\rm M}}=0.27$, $\Omega_{\Lambda}=0.73$,
 and $H_0=71\,\mbox{km}\,\mbox{s}^{-1}\,\mbox{Mpc}^{-1}$.
 All magnitudes mentioned are in AB mag, corrected for 
Galactic extinction using the
 Schlegel et al.(1998) dust map. 

\section{The Local Sample of H$\alpha$ Emitters}
 
  \subsection{Local H$\alpha$ Emitters}

 We selected objects with H$\alpha$ EW larger than 500\,$\mbox{\AA}$ 
 from Sloan Digital Sky Survey (SDSS) Data Release 7, using the MPA-JHU value added 
 catalog\footnote{http://www.mpa-garching.mpg.de/SDSS/DR7/} (Kauffmann et al. 2003a;
 Brinchmann et al. 2004). 
After the H$\alpha$ EW cut, 
we inspected each object individually in the SDSS image and spectrum. 
Extragalactic HII regions, which reside in the spiral arms of large $z\sim0$ galaxies,
are discarded. 
Due to the limited fiber size used in SDSS spectroscopy ($3\arcsec$ diameter), 
the measured H$\alpha$ EW corresponds to the equivalent width in the central regions of an extended
galaxy. 
Thus the galaxy-integrated H$\alpha$ EW 
may be different if the line to continuum ratio changes dramatically in the
outer parts of a galaxy.  
To ensure that gradients in equivalent widths do not significantly affect our results, 
we excluded objects with half-light radii larger than $3\arcsec$ 
(using $r_{50}$ in $r$-band; exponential disk fitting from SDSS pipeline).
Finally,
we are left with 299 galaxies ($\sim0.04$\%) out of 818333 galaxies 
spectroscopically observed in SDSS DR7. 
We refer these galaxies as `local HAEs'.

As discussed above, 
the fiber spectra only sample the central regions for extended objects. 
Since it is possible that the stellar continuum may be less
centrally concentrated compared to the line emission, aperture corrections 
are required as described in Brinchmann et al. (2004). 
Brinchmann et al.(2004) estimated the aperture correction 
of up to a factor of 10 to be applied to the 
star formation rates and stellar mass values in the MPA-JHU catalogs,
based on the likelihood distribution function P(SFR/$L_i$|color). 
In order to further ensure that the line flux measurements and equivalent width measurements are robust,
we convolved the SDSS spectrum with the bandpasses used for photometry, 
and compared the magnitudes derived in this way
with the observed broadband magnitudes. 
At redshifts greater than 0.05, the H$\alpha$ line falls within the Sloan $i-$band filter
while at lower redshifts, the line would be in the Sloan $r-$band filter. As a result,
aperture corrections based on either $r$ or $i$-band photometry 
would take any variation in equivalent width outside the fiber into account.
We found the median correction factor for the line flux estimates to be $\sim1.5$.

Table \ref{tab:objlist} lists the SDSS photo-object IDs, coordinates, 
H$\alpha$ flux 
and equivalent width, and other parameters for local HAEs. 
The H$\alpha$ flux and equivalent width are the observed values.
The analysis such as the derivation of H$\alpha$ luminosity
has been performed after applying aperture correction 
on the numbers presented in the Table \ref{tab:objlist}

\subsection{Ancillary Data Used}

Most parameters used in this analysis are drawn either from 
MPA-JHU value added catalog and/or the SDSS pipeline. 
Stellar mass, metallicity, line fluxes and equivalent widths for 
significant emission lines, 4000\,$\mbox{\AA}$ break $D_n$(4000) 
in the MPA-JHU value added catalog
are obtained through spectral and/or multi-wavelength photometry fitting. 
Here we briefly describe how each parameter is derived in the MPA-JHU value added catalogs.

Metallicities, i.e., the oxygen abundances, were derived
following the method presented in Tremonti et al.(2004).
All the most prominent emission lines
([OII], H$\beta$, [OIII], H$\alpha$, [NII], [SII])
were fit simultaneously to $\sim2\times10^5$ models for integrated galaxy spectra
(Charlot \& Longhetti 2001)
with varying metaliicity, ionization parameter, and dust-to-metal ratio.
The median of the likelihood distribution of the metallicity
was adopted as the metallicity of each galaxy.
 The stellar masses were derived based on the multi-wavelength photometry fitting,
following the method described in Kauffmann et al.(2003a) and Salim et al.(2007).
Template galaxy spectra were generated using Bruzual \& Charlot (2003)
population synthesis code, with 
metallicity range spanning $0.1-2\,Z_{\odot}$ and
 stellar population age spanning 0.1\,Gyr to the age of the universe.
An exponentially decaying star formation history was assumed
with random starbursts superimposed on the continuous star formation. 
Internal reddening and attenuation due to the IGM
are applied to the template galaxy spectra,
and the resulting template spectra were convolved with SDSS passbands to
produce broad-band photometry in each filter.
 After calculating $\chi^2$ of the fit, probability distributions for each
parameters, such as star formation rate, stellar masses, and dust attenuation
were constructed. Rather than finding single best-fit template,
the MPA-JHU value added catalog presents the median of the distribution of each parameter
in order to avoid the degeneracies between parameters.

  We have now added GALEX far- and near-ultraviolet (FUV/NUV) flux densities
of local HAEs from GALEX Data Release 6. GALEX has a spatial resolution of $\sim$6$\arcsec$.
 Using a cross-identification radius of 3$\arcsec$, 266 HAEs among 299 
($\sim90$\%) are observed in at least one GALEX band. After matching, each 
 result was visually examined to make sure that FUV or NUV emission is 
 from the local HAE itself, and not contaminated by nearby objects. 
 The flux density limit differs depending on which GALEX survey 
 the ultraviolet flux has been derived from - however on average, 
 the flux density limits in the FUV (1500\,$\mbox{\AA}$) and NUV (2300\,$\mbox{\AA}$)  
 are 5\,$\mu$Jy and 4\,$\mu$Jy respectively.

  We also obtained mid-infrared (MIR; $3-22\,\mu$m) photometry for local HAEs
 from the Wide-field Infrared Survey Explorer (WISE) preliminary data release (Wright et al. 2010).
 95 of the galaxies (95/299, $\sim30$\%) have coverage
 during the WISE preliminary survey period, 
 among which 65 galaxies are detected in either WISE channel 3 (12\,$\mu$m) 
 or channel 4 (22\,$\mu$m) with approximate flux density limits of
 $S_{12\mu m}>1$\,mJy and $S_{22\mu m}>4.3$\,mJy at the 5\,$\sigma$ level.  
 WISE channel 3 and channel 4 band fluxes are converted to 
 total infrared luminosities using bolometric corrections derived from model infrared galaxy templates 
 (Chary \& Pope 2010).

\section{Comparison with Other Galaxies}

    \subsection{Comparison with $z\sim4$ HAEs}

\begin{figure}
  \plotone{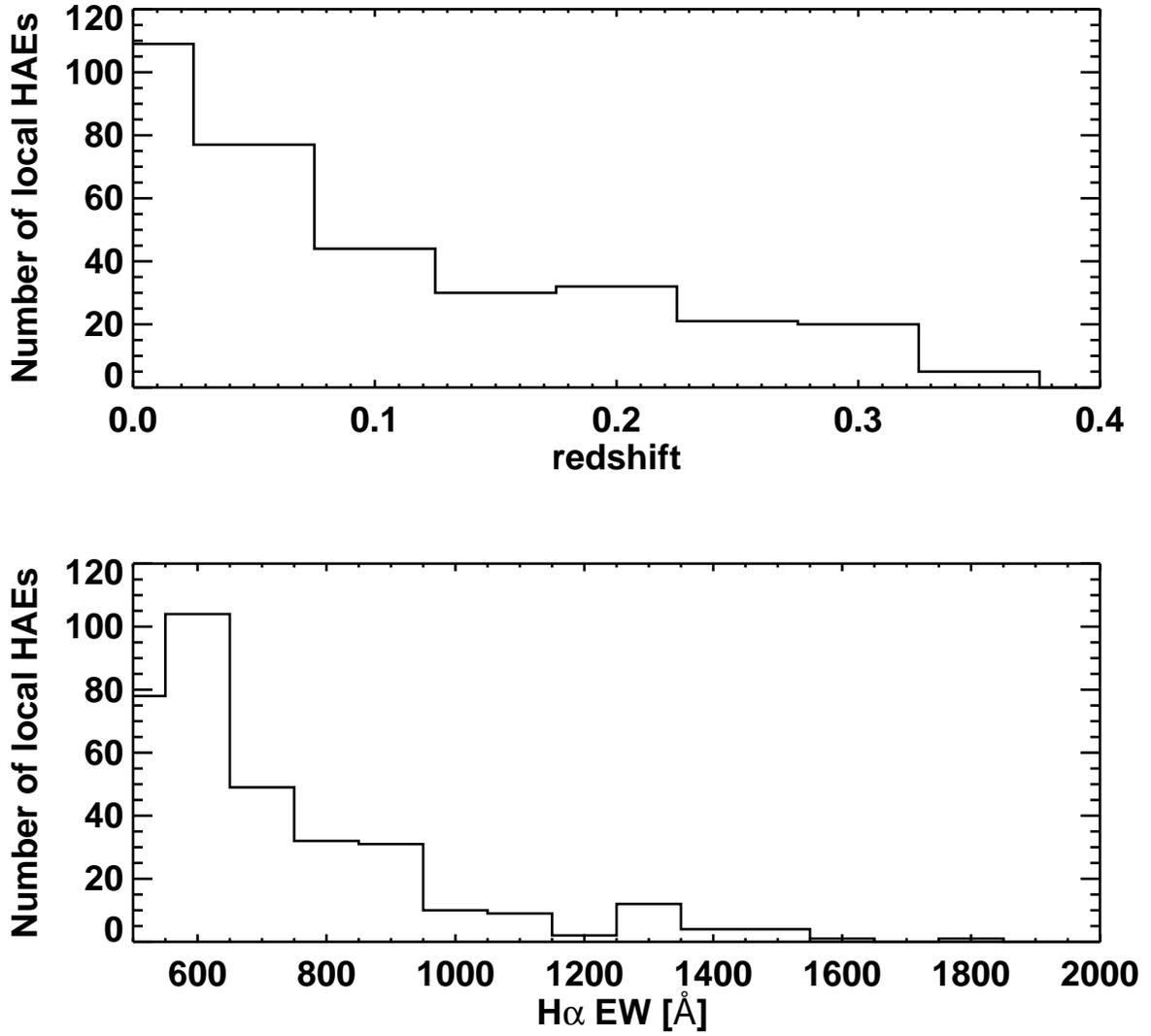}
   \caption{ \label{fig:EWzdist}
H$\alpha$ EW distribution (top) and redshift distribution (bottom) 
of 299 local HAEs. The EWs are the values in the observed frame, 
not in the rest frame.
   }
\end{figure}

\begin{figure}
\plottwo{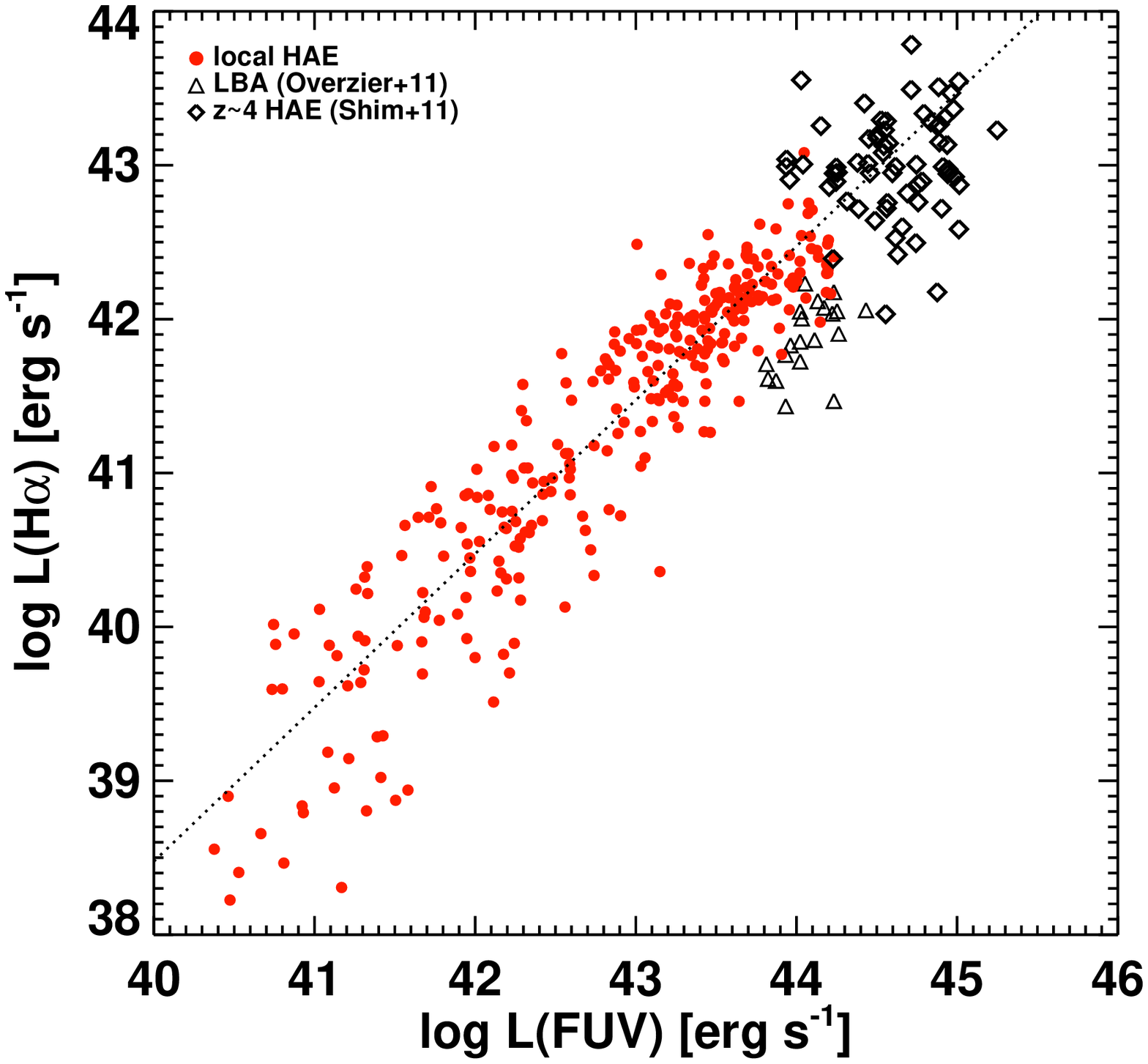}{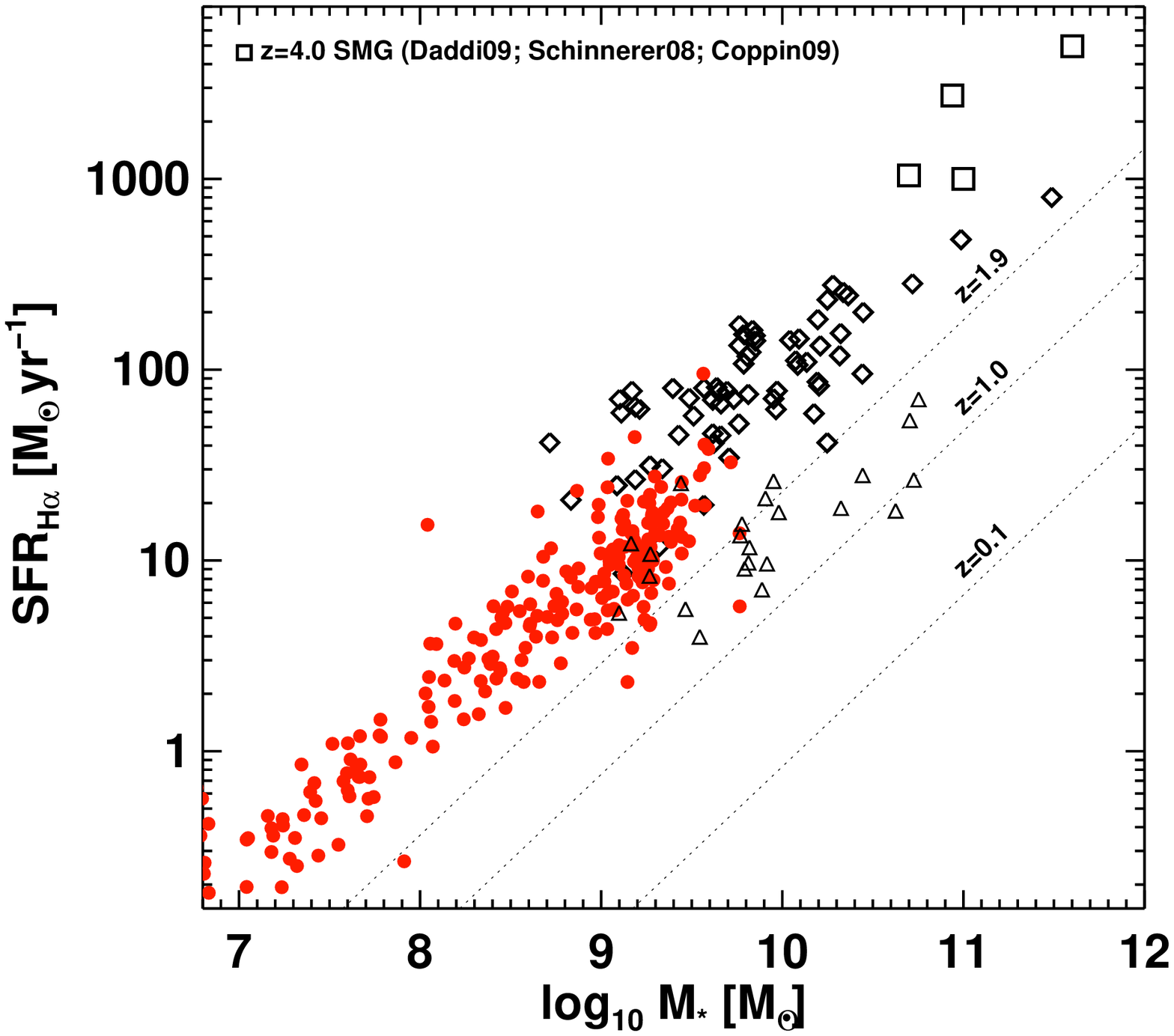}
  \caption{ \label{fig:compare_HAE_z4}
(\textit{Left}): Comparison between H$\alpha$ and FUV luminosity 
($\lambda L_{\lambda}$ at 1500\,$\mbox{\AA}$) 
of local HAEs (filled circles), local Lyman break analogs (triangles; Overzier et al. 2011)
and $z\sim4$ HAEs (Shim et al. 2011; diamonds). The
H$\alpha$ and UV luminosity of local HAEs is lower than that of $z\sim4$ HAEs 
by more than an order of magnitude, 
yet the H$\alpha$/UV luminosity ratio is consistent for both galaxy populations. 
The dotted line is a linear fit to describe L(H$\alpha$) vs. L(FUV) of local HAEs, 
and its slope value is almost unity (0.99). $z\sim4$ HAEs lie along this line, 
implying that $z\sim4$ HAEs are scaled-up versions of local HAEs. 
LBAs lie below the dotted line, suggesting that their H$\alpha$ luminosity 
is on average by a factor of 3 lower than HAEs with the same UV luminosity
(\textit{Right}): Comparison of SFR per unit stellar mass 
for local HAEs, LBAs and $z\sim4$ HAEs. 
The SFR is derived from the aperture-corrected (see Section 2.1) H$\alpha$ luminosity, 
using the relationship in Kennicutt (1998). 
As the overplotted dotted lines indicate,
star-forming galaxies appear to form a tight correlation. 
The SFR per unit stellar mass increases as the redshift increases
(Elbaz et al. 2007; Noeske et al. 2007; Daddi et al. 2007). 
$z\sim4$ HAEs produce SFR per unit stellar mass comparable
to $z=2.8-4.0$ submillimeter galaxies (Daddi et al. 2009; Schinnerer et al. 2008;
Coppin et al. 2009). 
Local HAEs show consistent SFR per unit stellar mass as $z\sim4$ HAEs
while LBAs lie well below the correlation,
suggesting that local HAEs are more robust analogs of $z\sim4$ star-forming galaxies 
with comparably high star formation efficiency. 
The high mass end of the stellar mass distribution of local HAEs overlaps 
with the low mass end for $z\sim4$ HAEs at $10^9-10^{10}\,M_{\odot}$. 
There are however observational selection
effects against identifying lower mass high-redshift HAEs.
}
 \end{figure}

 Local HAEs are distributed at $0<z<0.4$, with more than half lying at 
 $z<0.1$ (Figure \ref{fig:EWzdist}). Figure \ref{fig:compare_HAE_z4} shows how local HAEs and their
 $z\sim4$ counterparts (Shim et al. 2011) compare with each other in terms of 
the physical properties of their stellar populations. 
 In Figure \ref{fig:compare_HAE_z4}a, we compare the H$\alpha$ and FUV luminosity of
local HAEs, $z\sim4$ HAEs and LBAs from Overzier et al. (2011). The H$\alpha$/FUV luminosity
ratio of both local and $z\sim4$ HAEs are similar ($\sim0.032$)
while the local LBAs show
a ratio of $\sim0.0063$. 
The dotted line in Figure \ref{fig:compare_HAE_z4}a is a linear fit
to describe the correlation between H$\alpha$ luminosity and UV luminosity
of local HAEs, with a slope value of 0.99. Since both H$\alpha$ and UV are star-formation
tracers which are affected by age and dust obscuration in different ways, a slope close to unity indicates
that both sets of galaxies display similar physical properties of the stellar population and ISM. 
However, local HAEs have H$\alpha$ luminosities and UV luminosities lower than that of 
$z\sim4$ HAEs by nearly an order of magnitude, which means that the actual 
star formation rates of local HAEs are smaller than their high-redshift
counterparts by a factor of $\sim10$ (Figure \ref{fig:compare_HAE_z4}b). 
It is however clear that $z\sim4$ HAEs are scaled-up versions of local HAEs.  

 In Figure \ref{fig:compare_HAE_z4}b, 
we compare the star-formation rates relative to the stellar mass
of local HAEs, $z\sim4$ HAEs and LBAs. 
The H$\alpha$ based star-formation rates are derived assuming a 
Salpeter IMF and a constant star-formation rate.
The ratio of star-formation rate to stellar mass is referred to 
as the specific star-formation rate. Dotted diagonal lines in the figure show specific star-formation
rates at different redshifts of star-forming main sequence galaxies based on the literature 
(Elbaz et al. 2007; Noeske et al. 2007; Daddi et al. 2007).
Local HAEs show H$\alpha$ based star-formation rates that are almost 
a factor of 30 higher than typical star-forming
galaxies of the same stellar mass, i.e. 
they show elevated specific star-formation rates which are comparable to
that of $z\sim4$ HAEs and merger-driven high-redshift submillimeter galaxies. 
The stellar mass range sampled by local HAEs is on average lower than 
that of $z\sim4$ HAEs. 
 Although, there is a significant overlap at $10^{9} - 10^{10}\,M_{\odot}$;
 yet a dominant fraction ($\sim65$\,\%) of local HAEs appear to be small galaxies 
 with stellar mass smaller than $10^{9}\,M_{\odot}$.

 It is therefore clear that the H$\alpha$ EW $>500\mbox{\AA}$ selection
identifies local galaxies that
 are similar in their specific star-formation rates to large H$\alpha$ EW galaxies at $z\sim4$ but with scaled down
star-formation rates and stellar masses. 
Note that it is also possible that there could be $z\sim4$ HAEs 
with H$\alpha$ EW less than 500\,$\mbox{\AA}$. However, since our selection for
high redshift
HAEs is based on the broad-band photometric excess in the Spitzer 3.6\,$\mu$m imaging
data in the Great Observatories Origins Deep Survey fields (Shim et al. 2011) we are 
biased towards selecting the largest equivalent
width objects. Such large EW objects are also being found in wide-field spectroscopic
imaging surveys (Atek et al. 2011). Nevertheless, $\sim70$\% of spectroscopically selected 
$z\sim4$ galaxies (most of which are initially classified as Lyman break galaxies) 
are observed to show a photometric excess, reflecting strong H$\alpha$ emission 
with H$\alpha$ EW$>500\,\mbox{\AA}$.

     \subsection{Comparison with Local UV-selected Star-forming Galaxies}

 We next compare local HAEs with local analogs of z$\sim2$
 star-forming galaxies: UV luminous galaxies (UVLG) and Lyman break analogs 
 (LBAs) which are a subset of UVLGs. 

\begin{figure}
  \plotone{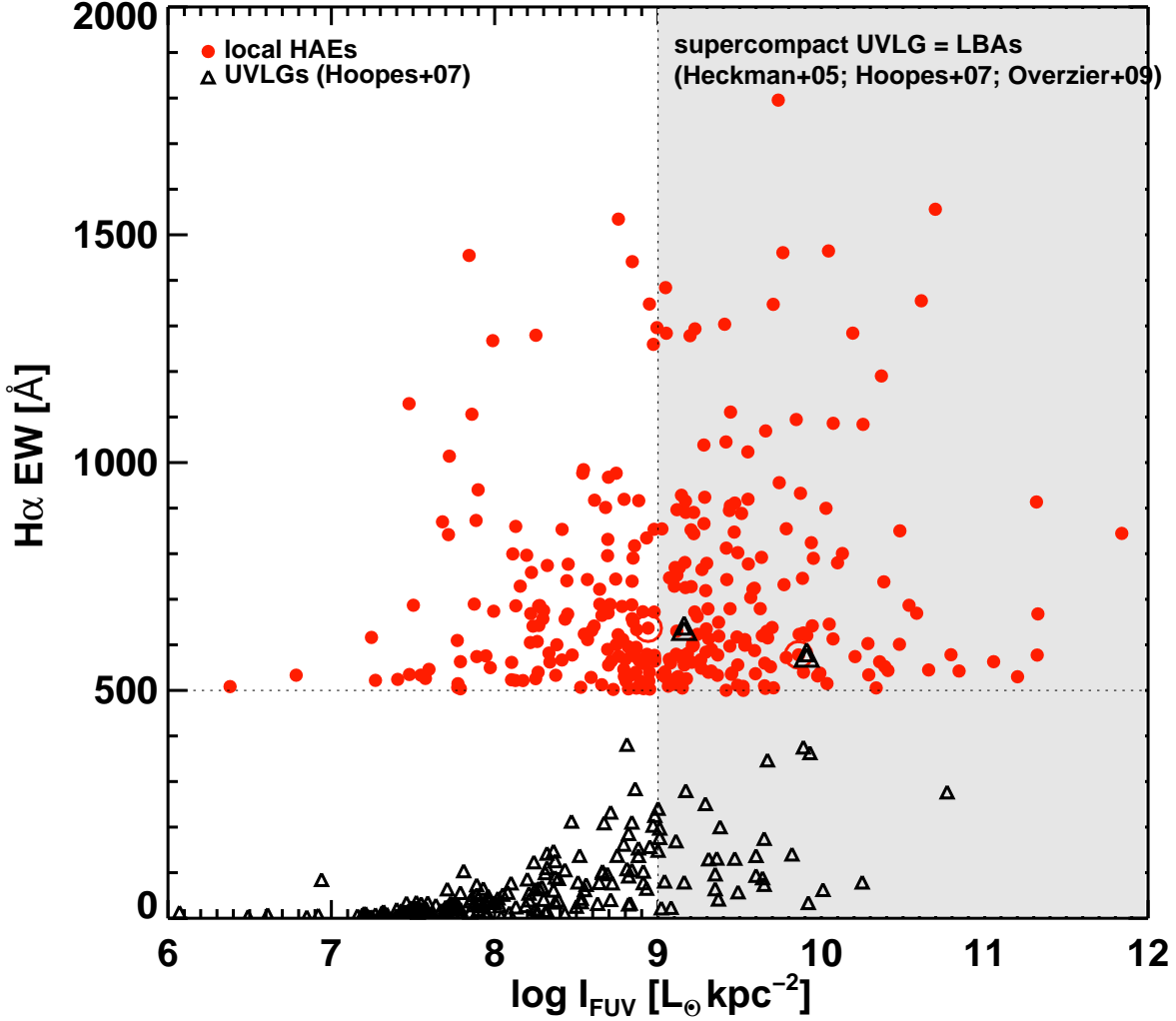}
  \caption{ \label{fig:compare_HAE_LBA}
Comparison between the UV surface luminosity 
and H$\alpha$ EW of local HAEs and LBAs. 
LBAs are a subset of UV-luminous galaxies (UVLGs, $L_{FUV}>10^{10.3}\,L_{\odot}$; 
Heckman et al. 2005; Hoopes et al. 2007) 
that have relatively high UV surface luminosity ($I_{FUV}>10^{9}\,L_{\odot}$\,kpc$^{-2}$). 
In this plot, local HAEs are plotted as filled circles, and 
UVLGs (Hoopes et al. 2007) are plotted as open triangles. Shaded region 
indicates the region where supercompact UVLGs, i.e., LBAs, fall. 
Although the total UV luminosity is not as luminous as that of LBAs, 
nearly 50\,\% of the local HAEs have as high a UV surface luminosity as that 
of LBAs.
Only two objects (symbols enclosed with larger symbols) 
are found to be overlapping between the local HAE and UVLG sample. 
}
 \end{figure}

 Contrary to the consistency between local HAEs and $z\sim4$ HAEs, 
 the local analogs of Lyman break galaxies (Heckman et al. 2005; Hoopes et al. 2007; 
Overzier et al. 2009) show significant differences with $z\sim4$ HAEs in terms of 
H$\alpha$ EW. Lyman break analogs (LBAs) share a similar stellar mass range 
with $z\sim4$ HAEs, but most of the LBAs have H$\alpha$ EW lower than 500\,$\mbox{\AA}$
(Figure \ref{fig:compare_HAE_LBA}). While LBAs have been shown to
have UV luminosity surface density consistent with that of Lyman break galaxies, 
their H$\alpha$ emission is significantly weaker than that of the $z\sim4$ Lyman break galaxies. 
This implies that either the star formation history and/or the physical parameters related 
to star formation are likely different between LBAs and high-redshift Lyman break galaxies. 
Furthermore, the LBAs are thought to be more analogous to the $z\sim2$
galaxy population where the H$\alpha$ EWs are correspondingly lower (Reddy et al. 2010, Erb et al. 2006).

In Figure \ref{fig:compare_HAE_z4}a,  
LBAs lie consistently below the dotted line describing the H$\alpha$-to-UV luminosity ratio 
for HAEs. On average, the H$\alpha$ luminosity of LBAs is $>3$ times lower than HAEs
with the same UV luminosity. Since H$\alpha$ EW is closely tied to the age of the stellar population
and the metallicity, it appears that the LBAs might have older, higher metallicity stellar populations
than both local and $z\sim4$ HAEs, a scenario we will investigate in greater detail in Section 5. Based on these observed differences, it appears that
the LBAs are unlikely to be representative of typical star-forming galaxies at $z\sim4$.

Figure \ref{fig:compare_HAE_LBA} illustrates the distribution of 
H$\alpha$ EW and UV surface brightness of local HAEs and UVLGs ($L_{FUV}>10^{10.3}\,L_{\odot}$).
The UV surface brightness is estimated at 1530\,$\mbox{\AA}$, i.e., the central wavelength of GALEX FUV-band.
 The surface brightness is calculated by dividing one-half the rest-frame 1530\,$\mbox{\AA}$ luminosity
 by the area of the galaxy encompassed within the half-light radius in the $u$-band
 ($I_{FUV}=L_{FUV}/2\pi r^2_{50,u}$; Hoopes et al. 2007).
 For the half-light radius in $u$-band, we used the radius from the seeing-corrected
 exponential model fit calculated by the SDSS pipeline
 which is consistent with the analysis of Hoopes et al.(2007). For galaxies
 larger than $2.2\arcsec$ which would be resolved even in seeing limited images, the observed half-light radius was used
 instead of the seeing-corrected radius.
 There are only two overlapping objects between the samples of local HAEs and UVLGs.
The SDSS object ids (photometric id) of these two are 587724199349387411 ($z=0.287$) 
and 588013384341913605 ($z=0.181$; Table \ref{tab:objlist}). 
 The two objects are marked as red points enclosed  
 with large red circles with $I_{FUV}$ values calculated using the technique 
outlined above.
 The triangles enclosed with larger triangles show the $I_{FUV}$ values for these two objects
in Hoopes et al. (2007), which are in excellent agreement with our values. 
 Discrepancies in the derived FUV surface brightness 
 originate from adopting different FUV magnitudes; 
 GALEX Data Release 2 (used in Hoopes et al. 2007) magnitudes
 are superceded by GALEX Data Release 6 (used in our analysis). 

 The median H$\alpha$ EW of UVLGs is only $\sim100\,\mbox{\AA}$,
 which is far less than the value used in the selection criteria of
 local HAEs ($>500\,\mbox{\AA}$). The LBAs, since they have higher UV surface brightnesses, 
should thereby show higher specific SFR (Hoopes et al. 2007), and thus larger H$\alpha$ EW. 
 The median H$\alpha$ EW of 
 compact ($I_{FUV}>10^9\,L_{\odot}$\,kpc$^{-2}$) UVLGs are indeed slightly larger than that of 
 large ($I_{FUV}<10^9\,L_{\odot}$\,kpc$^{-2}$) UVLGs (Figure \ref{fig:compare_HAE_LBA}),
 yet the correlation between UV surface brightness and H$\alpha$ EW of LBAs (compact UVLGs)
 is not significant. No such correlation can be seen in local HAEs.
 Remarkably, $\sim50$\,\%
 of local HAEs have UV surface brightness higher than $10^9\,L_{\odot}$\,kpc$^{-2}$, 
 comparable with that of LBAs.
 This high UV surface brightness is mostly 
 due to the small half-light radius of local HAEs rather than the UV luminosity itself.
 While UVLGs (thus LBAs) are defined as galaxies
 with $L_{FUV}$ larger than $2\times10^{10}\,L_{\odot}$ , 
 the $L_{FUV}$ distribution of local HAEs ranges between $5\times10^{6}-10^{11}\,L_{\odot}$. 

 Many local HAEs show compact morphologies 
 despite their redshift distribution skewed towards lower redshift compared to UVLGs.  
 The redshift distribution of UVLGs ranges between $z=0.05-0.3$, while the redshift
 distribution of HAEs ranges between $z=0.001-0.35$.
 As a result, the half-light radius of local HAEs is in general smaller than $5\,$kpc, 
 while the half-light radius ($r_{50,u}$) ranges between $1-15\,$kpc for UVLGs. 
 Because of this difference in $r_{50,u}$, the median UV surface brightness 
 ($\langle I_{FUV} \rangle$) of local HAEs ($10^{9}\,L_{\odot}$\,kpc$^{-2}$)
 is higher than that of UVLGs ($10^{8}\,L_{\odot}$\,kpc$^{-2}$).
This shows that there exists a significant number of local starbursts with specific star-formation rates
similar to that of high-redshift star-forming galaxies but which are missed in samples
that use the total UV luminosity as the primary selection criterion.

Despite the similarity in UV surface brightness,
the spectral features appearing in the composite spectrum of LBAs and local HAEs
are significantly different. 
 Figure \ref{fig:overlap} shows a comparison between the composite spectrum
of LBAs and local HAEs (top), as well as the comparison between the composite spectrum
of local HAEs and a single Wolf-Rayet galaxy, IC3591 (bottom).
The composite spectrum of local HAEs
is an average stack of the 197 local HAEs which do not show any signs of
Wolf-Rayet signatures such as the blue bump around HeII 4686\,$\mbox{\AA}$
and/or HeII 4686\,$\mbox{\AA}$ emission line with $S/N$ greater than 3. 
Before stacking, each spectrum was normalized using the continuum flux at 4800\,$\mbox{\AA}$.
The composite spectrum of LBAs 
is an average stacking of the 27 LBAs (Overzier et al. 2009) using the same method. 
 The composite spectrum of local HAEs shows slightly bluer stellar continuum 
at $4000-6000\,\mbox{\AA}$. When the optical stellar continuum is approximated using 
$F_{\lambda} \propto \lambda^{-\gamma}$, the slope $\gamma$ for local HAEs is 3.9 while
that for LBAs is 1.8, two times lower.\footnote{The slope is determined for arbitrary
$F_{\lambda}$ unit, thus the numbers should be used for relative comparison only.} 
The composite spectrum of local HAEs
also shows clear HeII 4686\,$\mbox{\AA}$ emission, 
 HeI 4471\,$\mbox{\AA}$ and 4921\,$\mbox{\AA}$ lines, and other forbidden lines 
 such as [FeIII] 4658\,$\mbox{\AA}$, [ArIV] 4711\,$\mbox{\AA}$ and 4740\,$\mbox{\AA}$ lines.
 The [NII] emission at either 6548\,$\mbox{\AA}$ or 6584\,$\mbox{\AA}$
 is relatively weak 
 compared to H$\alpha$, and [OIII] doublet at 5007\,$\mbox{\AA}$ is strong compared to 
 H$\beta$ in the local HAEs. 
The physical properties of the galaxies will be further discussed through the location of local HAEs in the
 emission line diagnostics diagram (Baldwin et al. 1981) in Section 4. 

     \subsection{Comparison with Wolf-Rayet Galaxies}
 
\begin{figure}
  \plotone{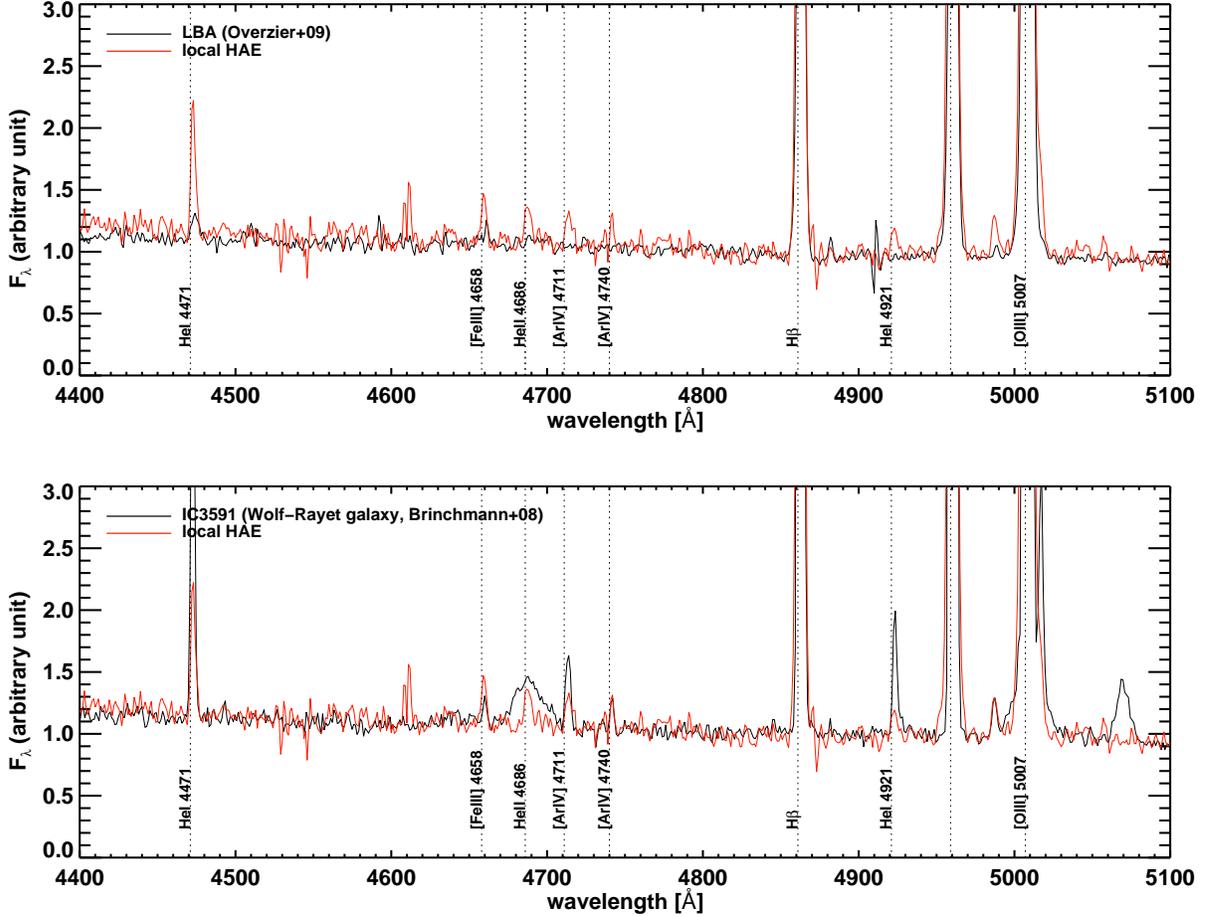}
  \caption{ \label{fig:overlap}
(\textit{Top}): Comparison between the composite spectrum of 197 local HAEs
that are not individually classified as Wolf-Rayet galaxies
and the composite spectrum of 27 LBAs (Overzier et al. 2009). 
When making a composite spectrum, each spectrum was weighted
using the continuum flux at 4800\,$\mbox{\AA}$ and no evidence of spectral features  
indicating the existence of Wolf-Rayet stars 
(such as blue bump around HeII 4686\,$\mbox{\AA}$ and/or HeII 4686\,$\mbox{\AA}$ line itself)
can be seen in each of the individual spectrum.. 
The composite spectrum of HAEs show 
HeII 4686\,$\mbox{\AA}$ line which is a distinguishing characteristic of Wolf-Rayet galaxies. 
Moreover, other Helium lines and  several metal lines at high ionization states
including [FeIII]\,4658\,$\mbox{\AA}$,
[ArIV]\,4771\,$\mbox{\AA}$, 4740\,$\mbox{\AA}$ appear in the composite spectrum of local HAEs. 
(\textit{Bottom}): Comparison between 
the composite spectrum of 197 local HAEs (same as above) and the spectrum of
a single Wolf-Rayet galaxy IC3591 (chosen from Brinchmann et al. 2008;
with H$\beta$ EW of 263\,$\mbox{\AA}$). 
The composite spectrum of local HAEs show several emission lines similar to
Wolf-Rayet galaxy, including He recombination lines and high-order Fe and Ar lines,
although the composite spectrum lacks the broad ``bump'' around 4686\,$\mbox{\AA}$.
However, the strength of emission lines appearing in this plot is more than 10 times lower 
in local HAEs than in Wolf-Rayet galaxies.
In addition, among 299 local HAEs, 43 galaxies (14\,\%) 
are individually identified as Wolf-Rayet galaxies
in previous works for Wolf-Rayet galaxies (Zhang et al. 2007; Brinchmann et al. 2008b).
This implies that the star forming environments in both the local HAE and Wolf-Rayet galaxies are similar.
   }
 \end{figure}

We matched the same of 299 local HAEs with the catalog of previously studied Wolf-Rayet galaxies
(Zhang et al. 2007; Brinchmann et al. 2008b).
In total, 43 local HAEs ($\sim14$\%) are identified as Wolf-Rayet galaxies.
Wolf-Rayet galaxies are specifically marked in Table \ref{tab:objlist}. 
Besides these 43, we found that another 59 local HAEs show 
HeII $\lambda4686$ emission line at line $S/N>3$ in their spectra,
despite they are not matched with Wolf-Rayet galaxy catalogs. 
For the remaining HAEs, we were able to construct a composite spectrum as described above. 
 At the bottom panel of Figure \ref{fig:overlap}, 
 we compare the composite spectrum of local HAEs (197 objects without 
Wolf-Rayet signature in individual spectrum; same as Figure \ref{fig:overlap}a)
with the spectrum of a Wolf-Rayet galaxy IC 3591 (Brinchmann et al. 2008b).
The composite spectrum does not show the existence of a broad blue bump typical of Wolf-Rayet galaxies. 
However, the fact that some of the local HAEs are Wolf-Rayet galaxies and the 
existence of the HeII 4686\,$\mbox{\AA}$ emission line in the composite spectrum
suggests that the
star formation environment in some local HAEs may be similar to that of Wolf-Rayet galaxies.

The origin of the HeII 4686\,$\mbox{\AA}$ line is however unclear and has been discussed extensively in
Thuan \& Izotov (2005) and Shirazi \& Brinchmann (2012). The former argue for radiative shocks while
the latter present a range of scenarios including a population of hot, early type stars due to stellar rotation.
The strength of HeII 4686\,$\mbox{\AA}$ is known to be dependent on H$\beta$ EW 
(Brinchmann et al. 2008b; Shirazi \& Brinchmann 2012).
 IC3591, compared here, has an H$\beta$ EW of 263\,$\mbox{\AA}$
(H$\alpha$ EW = 1556\,$\mbox{\AA}$) 
thus it is one of the galaxies with the largest H$\beta$ EWs among Wolf-Rayet galaxies. 
 We do not find any significant evidence for the strength of H$\alpha$ emission
in the local HAEs being a preferential identifier of Wolf-Rayet galaxies. 
However, it appears
that local HAEs are intermediate in their properties between Wolf-Rayet galaxies and LBAs in terms
of their emission line strengths as well as the strength of the intrinsic ionizing photon field which is
responsible for exciting the aforementioned lines. 
We discuss this in quantitative detail in the next section.

 \section{HeII Emission Line}

The HeII 1640\,$\mbox{\AA}$ emission line has long been suggested as 
a direct probe for hot Population III stars at high redshifts 
due to its high ionization potential of 54.4 eV (e.g. Schaerer 2003). 
Previous observations of high-redshift galaxies have only provided 
upper limits on HeII 1640\,$\mbox{\AA}$ 
(e.g. Dawson et al. 2007).
As illustrated in Figure \ref{fig:overlap}, 
the composite spectrum of local HAEs shows clear HeII 4686\,$\mbox{\AA}$. 
Moreover, some of the local HAEs (59 out of 305) show 
HeII 4686\,$\mbox{\AA}$ emission line over $S/N\sim3$ in their individual spectrum
even though they are not classified as Wolf-Rayet galaxies by previous studies 
(e.g., Brinchmann et al. 2008b). 
We measured the HeII 4686\,$\mbox{\AA}$ emission line flux
for local HAEs, and estimated the HeII 1640\,$\mbox{\AA}$ line flux 
using the intensity ratio between different transitions 
($n=4$ to $n=3$ for HeII\,$\lambda$4686, and $n=3$ to $n=2$ for HeII\,$\lambda$1640)
for case B recombination (e.g. Osterbrock \& Ferland 2006). 
We estimated the emission line flux by subtracting the stellar continuum through a simple linear fit
and fitting a single Gaussian to the residual emission.
 Except for the case of the stacked spectrum of the 197 local HAEs which were stacked,
we used line flux measurements from SDSS MPA-JHU value-added catalog for H$\alpha$
since the measurements take into account the effect of stellar absorption.
 
\begin{figure}
  \plotone{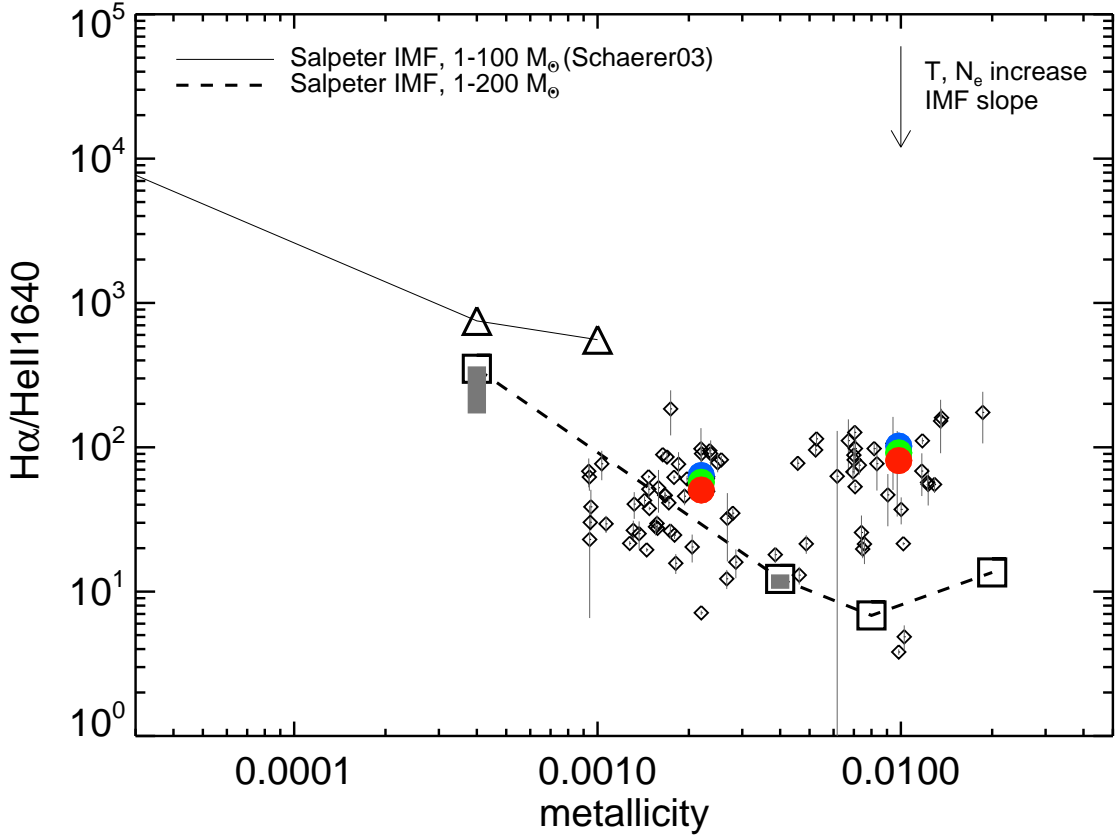}
  \caption{ \label{fig:HeII}
H$\alpha$-to-HeII 1640\,$\mbox{\AA}$ line flux ratio of local HAEs.
H$\alpha$ line flux is measured from the SDSS spectrum, while the
HeII 1640\,$\mbox{\AA}$ line flux is estimated using
the HeII 1640\,$\mbox{\AA}$/HeII 4686\,$\mbox{\AA}$ ratio (Osterbrock \& Ferland 2006)
and the measured HeII 4686\,$\mbox{\AA}$ line flux.
We measured the HeII 4686\,$\mbox{\AA}$ line flux using a simple Gaussian
line profile fit. Diamonds are local HAEs with HeII 4686\,$\mbox{\AA}$
detection in individual spectrum (including galaxies that are identified 
as Wolf-Rayet galaxies in Brinchmann et al. (2008b)). 
For local HAEs without individual HeII
detection, not classified as Wolf-Rayet galaxies, we stacked their spectra
and derived H$\alpha$/HeII ratio from the stacked spectrum (filled circles).
Circles with different colors show the possible range in H$\alpha$/HeII ratio
that can be spanned by changing the ionization temperature $T$ and/or
the electron density $N_e$.
Overplotted as dashed and solid lines are expected H$\alpha$/HeII ratio
from the STARBURST99 population synthesis models (Leitherer et al. 1999; Schaerer et al. 2003).
Thick gray rectangle shows the effect of changing the slope of the initial mass function
on the H$\alpha$/HeII line ratio; the slope has been changed from $-0.4$ (bottom)
to $-2.3$ (top).  The faint-end slope of the IMF,
mass limit of the IMF,
and the parameters regarding ionization environment ($T$, $N_e$)
have relatively small effects on the observed H$\alpha$/HeII ratio.
On the other hand, metallicity appears to be the most dominant factor
that affects the H$\alpha$/HeII line flux ratio.
}
\end{figure}

Figure \ref{fig:HeII} shows the ratio between H$\alpha$ flux and HeII\,$\lambda1640$ flux 
for the individual and stacked HAEs. 
We divided the 197 local HAEs into two bins of metallicity, based on the 
bimodal metallicity distribution illustrated in Figure \ref{fig:massmetal}. 
The size of each subsample is roughly the same ($\sim100$).
The color-coded filled circles represent the ratio from the stacked spectrum for two
subsamples, one at $Z\sim0.002$ and the other at $Z\sim0.01$. 
Different colors represent different ratios for the conversion 
from HeII\,$\lambda$4686 to HeII\,$\lambda1640$.
A ratio of HeII\,$\lambda$1640/HeII\,$\lambda$4686 of 
7.42 ($T=20000K$, $N_e=10^4$cm$^{-3}$) is shown in red,
6.79 ($T=10000K$, $N_e=10^6$cm$^{-3}$) in green,
6.56 ($T=10000K$, $N_e=10^4$cm$^{-3}$) in cyan,
and 5.90 ($T=5000K$, $N_e=10^4$cm$^{-3}$) in blue (Osterbrock \& Ferland 2006).  
Overplotted are the expected H$\alpha$/HeII\,$\lambda1640$ ratio from the Starburst99 population
synthesis model for a range of stellar initial mass functions and metallicity (Leitherer et al. 1999, Schaerer 2003). 

 As the different color-coded circles suggest, the observed H$\alpha$/HeII ratios
are not highly susceptible to variations in the physical parameters for HII regions 
 such as temperature and/or electron density. 
A change in temperatures and/or electron density 
produces only a 50\,\% 
change in the H$\alpha$/HeII ratio. 
 The effect of the initial mass function, including the mass range and the faint-end slope, is also limited to a factor of $2-3$. 
The observed H$\alpha$/HeII ratio for local HAEs is 
consistent with the expectation based on the Salpeter IMF with mass range of
$1-200\,M_{\odot}$ although the scatter in the points is large enough that a range of IMFs would be consistent with the data as shown by
the gray filled region in Figure \ref{fig:HeII}. 

The two most dominant factors that result in a wide range of 
H$\alpha$/HeII ratios are metallicity and star-formation history. 
As has been discussed in Schaerer (2003), metallicity strongly affects the ratio
of Hydrogen to Helium ionizing photons output due to the increase 
in the effective stellar temperatures with decreasing
metallicity. Thus, as metallicity decreases, the stellar population synthesis models 
result in a higher rate of Hydrogen
ionizing photons produced relative to the number of Helium ionizing photons, 
which translates to an increasing H$\alpha$/HeII\,$\lambda1640$ ratio. 
However, due to the short lifetimes of very massive, or hot Wolf-Rayet type stars 
that are responsible for the Helium ionizing photons,
after 5\,Myrs from the onset of star-formation, the HeII intensity drops 
while H$\alpha$ still remains elevated
due to the existence of late B and A stars. 
This implies that H$\alpha$/HeII\,$\lambda1640$ ratio increases at ages longer than 5\,Myr 
relative to ages below 5\,Myr.

Local HAEs span a range of metallicities as well as a range of HeII line strengths.
The direct detection of HeII in the majority of local HAEs as well as the stacked
detection of HeII in the HAEs which do not show HeII in their individual spectra
implies that the HAEs harbor a very young stellar population. It is challenging
to constrain the age of the stellar population very precisely due to the absence of 
obvious age indicators such as the 4000\,$\mbox{\AA}$ break (See Section 5.2). 
For population synthesis models with continuous star-formation, 
the H$\alpha$/HeII\,$\lambda1640$ ratio
would be $\sim$12. Since $>$90\,\% of local HAEs are above that ratio, the implied age
of the stellar population in these galaxies must be less than 10\,Myr strongly suggestive
of local HAEs being powered by short bursts. The morphologies of these galaxies are however
ambiguous in seeing-limited ground-based Sloan data. Space based data is required to assess if 
obvious signs of mergers are present in these objects.

 Since we have demonstrated that local HAEs are analogs of $z\sim4$ HAEs, we can estimate
the HeII\,$\lambda1640$ emission line flux from the H$\alpha$ line flux of $z\sim4$ HAEs,
and assess the feasibility of detecting the HeII\,$\lambda1640$ emission line in the spectra of 
high-redshift sources. 
Using the H$\alpha$/HeII ratio of $\sim50$ (values measured in the stacked spectrum
of low-metallicity local HAEs) and the median H$\alpha$ luminosity of
$10^{43}$ erg\,s$^{-1}$ (Shim et al. 2011), the
HeII\,$\lambda1640$ line flux of $z\sim4$ HAEs is expected to be 
$1.6\times10^{-18}$\,erg\,s$^{-1}$\,cm$^{-2}$.  
This is comparable but slightly lower than the HeII upper limits of Ly$\alpha$ emitters
at $z\sim4$ ($<$2$\times10^{-18}$\,erg\,s$^{-1}$\,cm$^{-2}$; Dawson et al. 2007). 
However, it should be noted that the relationship 
between Ly$\alpha$ emitters and H$\alpha$ emitters is currently ambiguous. 
The H$\alpha$ emitters in Shim et al. (2011) are 
typical spectroscopically confirmed LBGs while the LAEs are preferentially 
lower mass systems. Therefore to detect the HeII\,$\lambda1640$
emission line in $z>4$ HAEs, future observations should aim
to achieve line flux limits down to $5\sigma\sim10^{-18}$\,erg\,s$^{-1}$\,cm$^{-2}$.
 
 \section{Properties of Local HAEs: Origins of Strong H$\alpha$ Emission}

In Shim et al. (2011),
we presented several possible reasons for the origin of unusually strong H$\alpha$ emission 
in the $z\sim4$ HAEs,
 especially compared to other star-formation tracers such as the UV continuum. 
 The possible factors that drive strong H$\alpha$ emission in high-z HAEs are:
 (1) extinction properties which are different with that of
 local starbursts (i.e., extinction curve steeper in UV), 
 (2) young stellar population ages and star formation history, 
 (3) low metallicity, and 
 (4) a stellar initial mass function with an overabundance of massive stars.
AGN were ruled out as a plausible explanation based on the non-detection of the $z\sim4$ HAEs in the deep {\it Chandra} X-ray data.
In the previous section, we had investigated the effect of stellar initial mass function. 
 In this section, we investigate each of these possible scenarios for the origin of strong H$\alpha$ emission
 in local HAEs using the diagnostic properties of SDSS spectroscopy and UV-to-MIR photometry. 

\subsection{Low Metallicity}

\begin{figure}
 \plotone{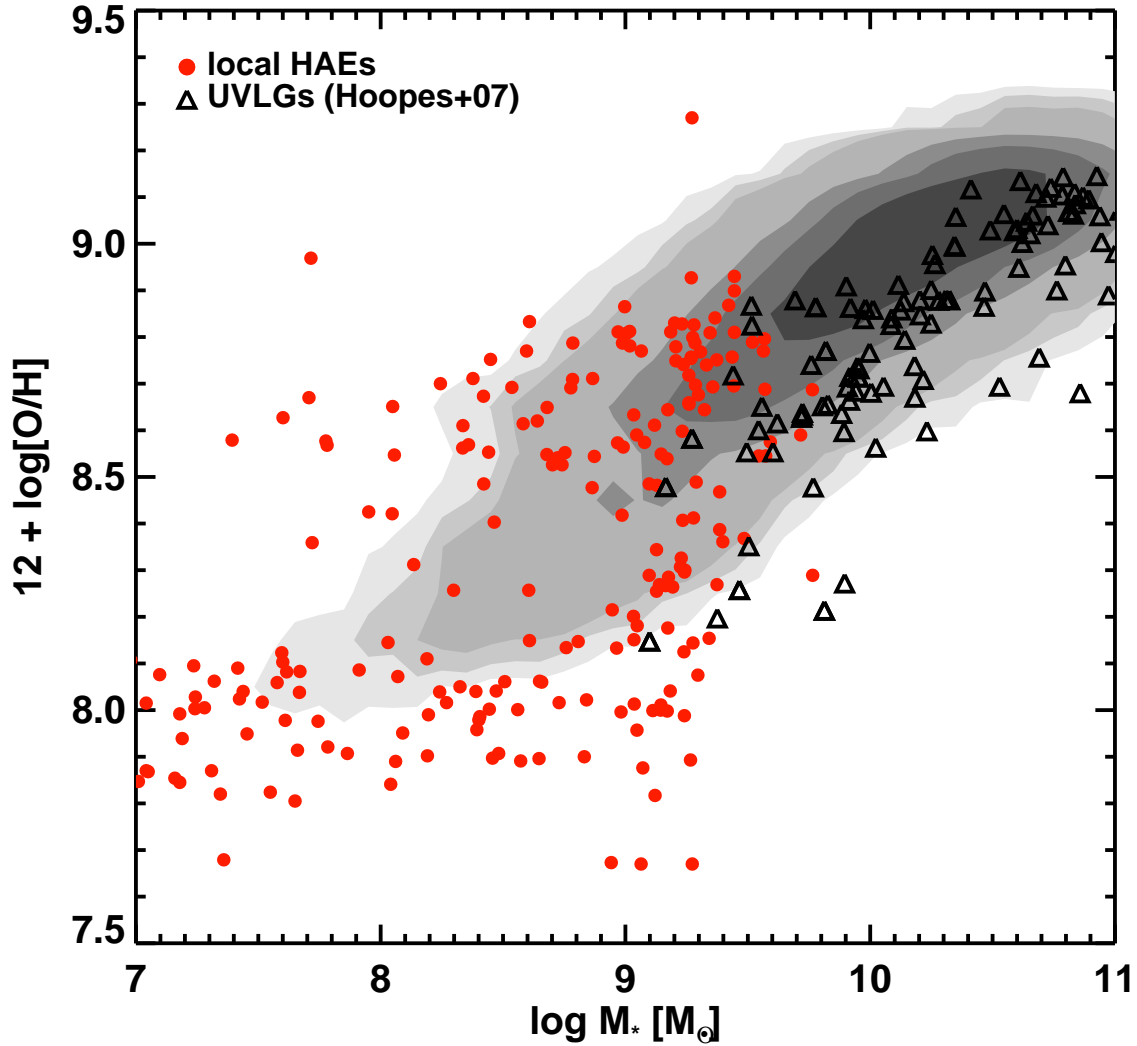}
 \caption{ \label{fig:massmetal}
Mass-metallicity relation of the local HAEs (filled circles). 
Gas-phase metallicity and stellar mass for local HAEs are from 
the MPA-JHU value-added catalog constructed using SDSS DR7.
 Stellar mass has been derived based on photometry fits 
(methods from Kauffmann et al. 2003a; Salim et al. 2007)
and gas-phase metallicity based on the emission line modeling 
(Tremonti et al. 2004; Brinchmann et al. 2004).
Local HAEs are plotted over the contours which show the density distribution
of entire SDSS DR7 galaxies that have been spectroscopically observed. 
Also compared here are UVLGs (triangles; Hoopes et al. 2007; $L_{FUV}>10^{10.3} L_{\odot}$), 
among which the most compact galaxies are classified as LBAs. 
While half of the local HAEs are considered to be at the lowest end of metallicity distribution, 
most of the UVLGs are more metal-rich than local HAEs. In addition to metallicity, 
UVLGs are on average more massive than local HAEs. 
The difference between the two populations suggests that LBAs are possibly a ``more evolved'' galaxy population 
compared to HAEs.
}
\end{figure}

Figure \ref{fig:massmetal} shows
gas-phase metallicities ($12+\mbox{log}[O/H]$) and stellar masses of local HAEs
from the MPA-JHU value added catalog for SDSS DR7.
 The metallicity distribution of local HAEs is clearly bimodal,
 one peak at $12+\mbox{log[O/H]}\sim8.6$ and another peak at $12+\mbox{log[O/H]}\sim8.0$.
 Half of the local HAEs are located at the low end of 
 the stellar mass distribution ($M_*<10^9 M_{\odot}$)  
 and the low end of the metallicity distribution ($12+\mbox{log[O/H]}<8.3$). 
This is different than the case for UVLGs (Hoopes et al. 2007).
Almost all UVLGs are more massive than $10^9 M_{\odot}$ and have metallicity
$12+\mbox{log[O/H]}>8.3$. 
The most metal-rich UVLGs are almost super-solar metallicity. 
The higher metallicity of UVLGs, the higher masses compared to local HAEs 
and the absence of HeII in the spectra of UVLGs
suggests that UVLGs are likely to be at a later stage of their star-formation history
compared to HAEs. 

 The [OIII]/H$\alpha$ ratios of local HAEs do not appear to 
correlate with H$\alpha$ equivalent widths with a median line
flux ratio of [OIII]/H$\alpha\sim1.7$. Thus it is difficult to assess
the connection between the gas-phase metallicity and 
the H$\alpha$ equivalent width. However, there is a clear trend
where the lower metallicity systems appear to show
a factor of 2 weaker [OIII] line flux corresponding to their H$\alpha$ 
line flux than the highest metallicity systems in the sample.

\begin{figure}
 \plotone{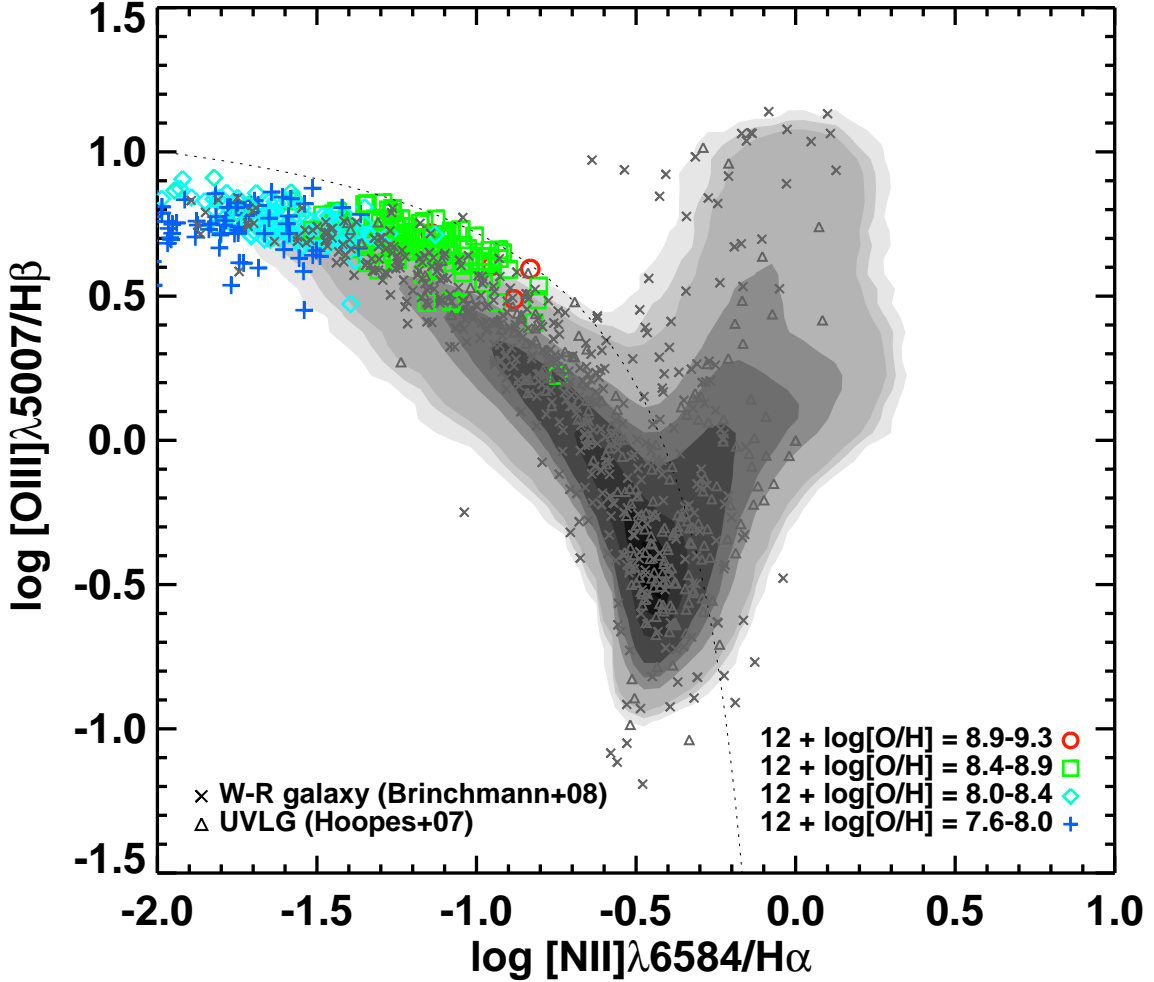}
 \caption{ \label{fig:BPT}
BPT diagram (Baldwin et al. 1981) of local HAEs. 
Contours show the density distribution of all emission line galaxies in the SDSS DR7. 
The dotted line represents the criterion that divides AGN-dominated galaxies (upper right)
and star formation-dominated galaxies (lower left) as defined in Kauffmann et al.(2003b). 
 Local HAEs with different gas-phase metallicities are plotted using different symbols:
red circles correspond to most metal-rich galaxies ($8.9<12+\mbox{log[O/H]}<9.3$)
and the metallicity decreases as the symbol change to 
green squares, cyan diamonds, and blue crosses.
 It is clear that local HAEs are not AGN-dominated systems, 
especially due to their weak [NII]\,6584\,$\mbox{\AA}$ emission. 
The [NII]\,6584\,$\mbox{\AA}$/H$\alpha$ ratio is proportional to 
the gas-phase metallicity, thus the weak [NII] of local HAEs indicate
that these are metal-poor systems.  Also shown are the locations of 
Wolf-Rayet galaxies (Xs; Brinchmann et al. 2008) 
and UVLG-LBAs (triangles; Hoopes et al. 2007) in the BPT diagram. 
Unlike the HAEs, Wolf-Rayet galaxies and the UVLGs 
do contain AGN dominated systems. 
}
\end{figure}

The low metallicity of local HAEs is also reflected 
in their [NII]$\lambda$6584/H$\alpha$ ratios, i.e.,  
 local HAEs show weak [NII] emission compared to H$\alpha$. 
The median value of the observed [NII]$\lambda$6584/H$\alpha$ is $\sim0.08$, 
far less than the conventional [NII]$\lambda$6584/H$\alpha=0.3$
for local star-forming galaxies (Gallego 1997). 
This places local HAEs well to the left of the line 
that divides starbursts and active galactic nuclei in the BPT diagram (Baldwin et al. 1981;
Kauffmann et al. 2003b; Kewley et al. 2001), 
eliminating the possibility of the strong H$\alpha$ emission arising due to the presence of AGN. 
Figure \ref{fig:BPT} shows the location of local HAEs in the 
[OIII]5007/H$\beta$ vs. [NII]6584/H$\alpha$ diagram.  
he location of the local HAE population in the BPT diagram
partly depends on the gas-phase metallicity ($12+\mbox{log[O/H]}$),
i.e., galaxies with lower metallicity show even less [NII]/H$\alpha$ ratio
than galaxies with higher metallicity
(the different symbols of different colors trace the metallicity).  
 Also shown for comparison are the location of Wolf-Rayet galaxies (Brinchmann et al. 2008b)
 and UVLGs (Hoopes et al. 2007) in the BPT diagram. 
 Significant numbers of either Wolf-Rayet galaxies or UVLGs are
 classified as AGN-dominated systems, 
 while most of Wolf-Rayet galaxies and/or UVLGs show
 higher [NII]/H$\alpha$ compared to local HAEs. 
 Again as in Figure \ref{fig:massmetal},  
 the BPT diagram suggests that UVLGs, as well as Wolf-Rayet galaxies in general,
 are on average more evolved than local HAEs. 

 Another interesting point in Figure \ref{fig:BPT} is 
 that local HAEs lie at the highest boundary of the entire local galaxy population
 showing the highest [OIII]/H$\beta$ ratios. 
 It means that there exists a non-negligible offset between 
 the median ridge line of [OIII]/H$\beta$ ratio for local galaxies
 and that of local HAEs ($\sim$1.6 in [OIII]/H$\beta$ line flux;
Kewley \& Dopita 2002).
 This offset is also reported for star-forming galaxies at $z>2$
 (Erb et al. 2006). 
 Brinchmann et al.(2008a) suggest that this offset from the BPT ridge line 
 is closely related to the amount of star formation, 
 it being roughly proportional to the H$\alpha$ EW. 
High-redshift galaxies lying well above the ridge line 
are producing unusually large amounts of stars compared to their stellar mass, 
i.e., showing larger specific SFR, 
and this displacement can be achieved by increasing the ionization parameter 
(the ratio of the volume densities of ionizing photons and particles). 
Again, the increase of the ionization parameter 
depends on the ratio of ionizing photons, electron densities, and 
the geometry of HII regions which determines whether the HII regions are 
density-bound or ionization-bound:
Brinchmann et al.(2008a) suggest higher electron densities
and larger escape fraction of hydrogen ionizing photons as
two major reasons for this offset in the BPT diagram. 
These can be also applicable to local HAEs,  
considering that local HAEs show a clear displacement in the BPT diagram comparable to
other $z>2$ star-forming galaxies. 
Thus, local HAEs appear to be excellent laboratories
for the study of high-z star-forming environments and a measure of the ionizing photon flux
from these objects would validate the origin of the unusual [OIII]/H$\beta$ ratios.

 \subsection{Star Formation History}

In Shim et al.(2011), we demonstrated that the most dominant factor
 that drives the `strong H$\alpha$ phase' is the star formation history of
 high-z star-forming galaxies. The ubiquity of strong H$\alpha$ emission
and the evolved stellar population ages indicate that $z\sim4$ star-forming galaxies
 appear to display continuous star formation histories rather than burst-like star formation.
 At $z\sim4$, the interpretation
 corroborates a star formation mechanism that is powered by a continuous gas supply
such as cold gas accretion from the filaments into massive halos.
 Short timescale events such as mergers would be stochastic and would 
suggest a strong H$\alpha$ emitting phase for only 10\% of the galaxies.

\begin{figure}
 \plotone{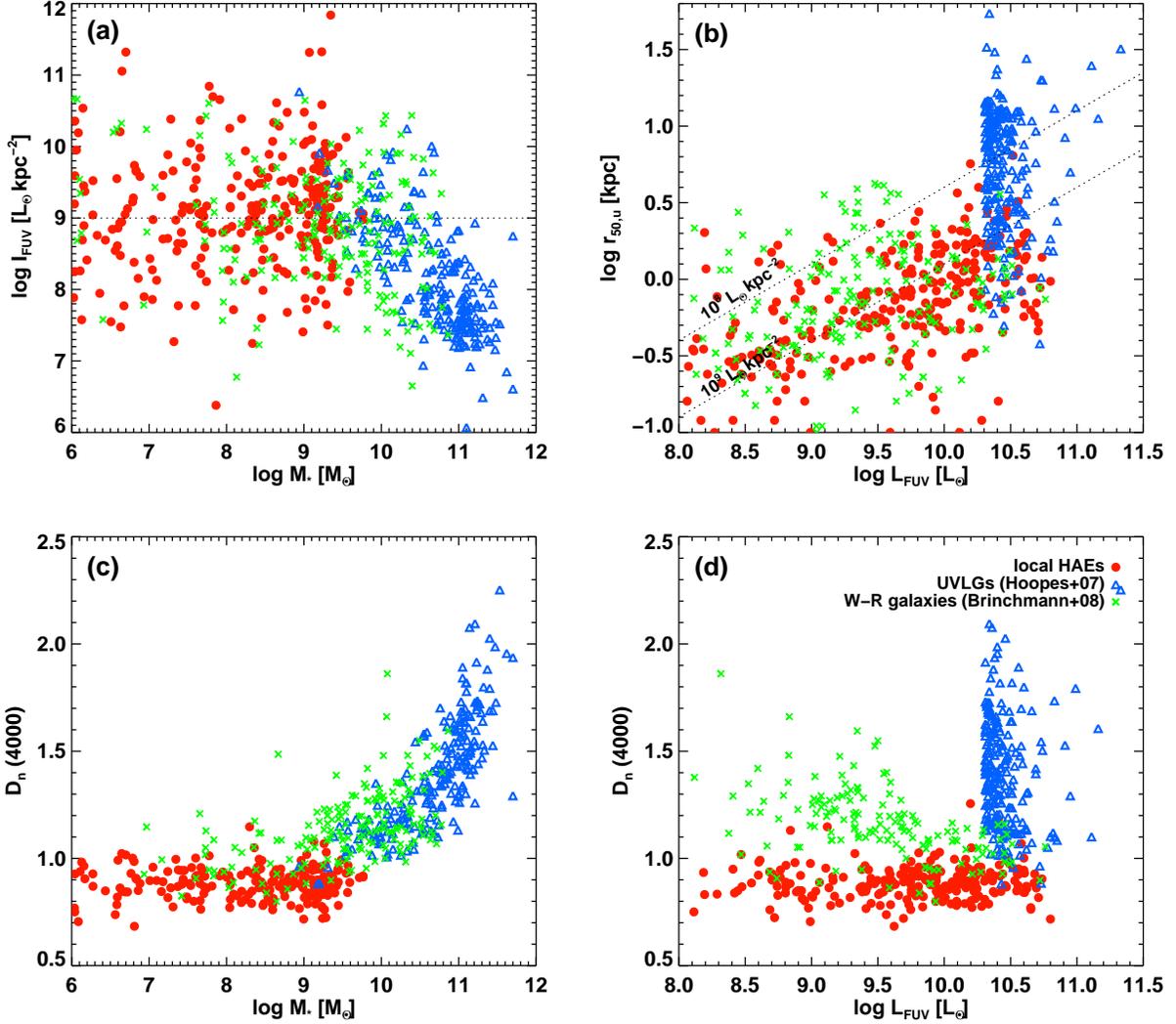}
  \caption{ \label{fig:Dn}
\small{(a) FUV surface luminosity density (at 1500\,$\mbox{\AA}$) vs. stellar mass for
local HAEs (filled circles), UVLGs (triangles; Hoopes et al. 2007),
and Wolf-Rayet galaxies (crosses; Brinchmann et al. 2008b).
 Dotted horizontal line at $I_{FUV}=10^{9}\,L_{\odot}$\,kpc$^{-2}$
indicates the criteria with which compact UVLG (LBA; Overzier et al. 2009)
and non-compact UVLG are divided.
UVLGs are the most massive among the three populations, 
although the FUV surface luminosity of HAEs and Wolf-Rayet galaxies are as high as
that of UVLGs. 
(b) Half-light radius in SDSS $u$-band ($r_{50, u}$) and FUV luminosity of
local HAEs, UVLGs, and Wolf-Rayet galaxies.
Dotted lines indicate FUV surface luminosity density of $10^{8}, 10^{9}$\,L$_{\odot}$\,kpc$^{-2}$
($y$-axis in Figure \ref{fig:Dn}a).
Since the half-light radius is determined through exponential disk-fitting
on the seeing-limited Sloan images, some `unresolved' local HAEs
are interpreted to have unreliably small half-light radius which are
less than 100\,pc.
(c) 4000\,$\mbox{\AA}$ break strengths ($D_n\,(4000)$) and stellar mass
of local HAEs, UVLGs, and Wolf-Rayet galaxies.
It is clear that UVLGs are more evolved galaxy populations than HAEs, 
while HAEs are considered to be much younger from their low $D_n\,(4000)$. 
(d) 4000\,$\mbox{\AA}$ break strengths and FUV luminosity
of local HAEs, UVLGs, and Wolf-Rayet galaxies.
Wolf-Rayet galaxies and HAEs 
show similar FUV luminosity, galaxy size,
and thus FUV surface luminosity which is also consistent with that of 
high-redshift star-forming galaxies.
This suggests that the environment of star forming regions of these two galaxy populations 
is consistent with each other. 
However, two galaxy populations show differences in $D_n\,(4000)$ and stellar mass. 
 The $D_n\,(4000)$ of Wolf-Rayet galaxies is not as homogeneously small as HAEs  and 
HAEs are slightly less massive than Wolf-Rayet galaxies.
Therefore, it is possible that there exists an evolutionary link between
two population: Wolf-Rayet galaxies and HAEs share the same star formation
environments, while HAEs are earlier stage of Wolf-Rayet galaxies 
before the evolution of metallicity and stars that enhance the 4000\,$\mbox{\AA}$ break.  }
}
 \end{figure}

In order to study the ages of the stellar population in local HAEs
we compare the strength of the 4000\,$\mbox{\AA}$ break, 
$D_n\,(4000)$, of these galaxies and UVLGs
in Figure \ref{fig:Dn}. 
Local HAEs sample a broad range of stellar masses covering 
$10^{6}-10^{10} M_{\odot}$,
a broad range of FUV luminosity covering 
$10^{8}-10^{11} L_{\odot}$,
yet in terms of $D_n\,(4000)$,
local HAEs are a relatively homogeneous population
with $D_n\,(4000) < 1.0$. 
 There is no correlation between $D_n\,(4000)$ and stellar mass or FUV luminosity. 
On the other hand, there is a positive correlation between
$D_n\,(4000)$ and the stellar mass of UVLGs: 
older UVLGs appear to be more massive. 
UVLGs also appear to be in the middle of the range in the $D_n\,(4000)$-M$_{*}$ properties
compared to Wolf-Rayet galaxies and HAEs.

Wolf-Rayet galaxies appear to show properties between those of HAEs and UVLGs.
The $D_n\,(4000)$ of Wolf-Rayet galaxies is not as homogeneously small as HAEs. 
Furthermore, Wolf-Rayet galaxies are more massive than HAEs but less massive than UVLGs. Wolf-Rayet galaxies therefore
appear to be older than HAEs of the same UV luminosity based on the
strength of their 4000\AA break but are younger than the majority of UVLGs. 
 The composite spectrum of HAEs do show emission lines
 observed in Wolf-Rayet galaxies, such as Helium recombination lines which suggests some similarity in their
 ionizing photon field. 
 The consistent FUV luminosity (Figure \ref{fig:Dn}a), 
 galaxy size (Figure \ref{fig:Dn}b),
and thus FUV surface brightness between the two 
also suggests the hypothesis that Wolf-Rayet galaxies and HAEs 
share similar interstellar medium condition, 
that enables high ionization parameter 
-- higher electron densities, higher temperature, 
and a large escape fraction for hydrogen ionizing photons.
Since the $D_n\,(4000)$ features arises due to absorption lines
from ionized metals, the difference in $D_n\,(4000)$ between two populations
suggests the possibility that
HAEs are at an earlier lower metallicity stage of evolution than Wolf-Rayet galaxies.

\subsection{Dust Obscuration in Local HAEs} 
 
In this subsection, we present and compare several different extinction indicators of local HAEs,
constrain their extinction properties and thereby their true star-formation rate.
We also assess the possible implications 
for the shape of the differential extinction as a function of wavelength. 

 \subsubsection{Extinction Indicators}

The first observable extinction indicator is the UV spectral slope $\beta$.
$\beta$ is derived based on the assumption that the
spectrum of star-forming galaxies in the UV wavelength range 
are well described by a power-law ($f_{\lambda}\propto{\lambda}^{\beta}$). 
We assumed that this power-law approximation 
is applicable to the rest-frame wavelength range of 
$1500/(1+z)\mbox{\AA}-2300/(1+z)\mbox{\AA}$, covered by the GALEX FUV and NUV bands.
Then we converted the $(FUV-NUV)$ color to a spectral slope $\beta$ using the following relation. 

 \begin{equation}
  \beta_{FUV-NUV} = (m_{FUV} - m_{NUV})/0.464 - 2.0
 \end{equation}

$m_{FUV}$ and $m_{NUV}$ indicate the GALEX magnitudes in the FUV- and NUV-band
corrected for Galactic extinction.
The low $S/N$ of the local HAEs in the GALEX FUV and NUV bands propagates into
the derived $\beta$. The derived $\beta_{FUV-NUV}$ is highly uncertain
in most cases; 40\,\% of all objects have uncertainty higher than $0.2$ in $\beta$.

 The above assumption is accurate for objects with $z<0.1$. However,
for objects at $z>0.1$, the Ly$\alpha$ emission line may contaminate 
the observed flux density in the GALEX FUV band 
which makes the broadband colors bluer and induces some uncertainty in $\beta$. 
For standard Case B recombination, the Ly$\alpha$ to H$\alpha$ line ratio is a factor of $\sim$10.
We estimate that in the most extreme case, this would bias the FUV flux blueward by 20\% and thereby
affect the UV slope $\beta$ by 0.4. We have not applied this correction because we do not know the true
strength of the Ly$\alpha$ line in these sources and due to the fact that the effect of GALEX photometric
uncertainty itself on $\beta$ values is typically larger. Furthermore, the comparison with UVLGs and LBAs
is more straightforward since those studies chose not to apply the correction as well.

The next extinction indicator is the flux ratio between 
Hydrogen recombination lines that traces 
the different extinction at the wavelengths corresponding to the lines. 
Here, we present the Balmer line ratio $F_{H\alpha}/F_{H\beta}$
as the extinction indicator. 
Following Calzetti et al. (2000), the color excess of the nebular gas, $E(B-V)_{gas}$, is calculated  
by comparing the observed Balmer line ratios 
with the intrinsic (i.e., unobscured) Balmer line ratios. 

 \begin{equation}
   E(B-V)_{gas} = \frac{2.5}{k(\lambda_1)-k(\lambda_2)} \mbox{log}_{10} \frac{F_i^{\lambda_1}/F_i^{\lambda_2}}{F_o^{\lambda_1}/{F_o^{\lambda_2}}}
 \end{equation}

Here, $F_i^{\lambda_1}/F_i^{\lambda_2}$ 
is the intrinsic line ratio,
$F_o^{\lambda_1}/{F_o^{\lambda_2}}$ is the observed line ratio,
and $k(\lambda)$ is an extinction value at the corresponding wavelength 
described by the applied extinction curve. 
Based on the case-B recombination at $T=10^4$\,K (Osterbrock \& Ferland 2006),
the intrinsic Balmer line ratio 
is $F_{H\alpha}/F_{H\beta}=2.87$. 

Finally, the ratio between IR luminosity and UV luminosity can be used
as a third extinction indicator, 
assuming that the IR luminosity represents the entire unobscured stellar radiation.   
We derived IR luminosities of local HAEs using the WISE photometry
in channel 3 and 4 (12$\mu$m and 22$\mu$m respectively).
By matching the coordinates of local HAEs to the WISE preliminary data release catalog
(Wright et al. 2010), we found 65 objects with robust counterparts 
at a flux density level higher than 0.28\,mJy at 12\,$\mu$m and 1.9\,mJy at 22\,$\mu$m. 
These objects do not have nearby neighbors within $3\arcsec$ 
that could possibly contaminate the MIR flux density. 
We used IR spectral energy distribution templates of IR galaxies with different 
IR luminosities (Chary \& Pope 2010): each IR luminosity template was redshifted
to the corresponding redshift of the object 
and the template that best explains the observed 12\,$\mu$m and 22\,$\mu$m flux density
was determined, while the IR luminosity of the object was derived to be the 
IR luminosity associated with the best-fit template. 
IR luminosity of local HAEs ranges between 
$1.4\times10^{8}\,L_{\odot}$ and $7.6\times10^{11}\,L_{\odot}$. 
We compared $L_{IR}$ with $L_{UV}$, 
i.e., $\lambda L_{\lambda}$ at 1530\,$\mbox{\AA}$ as defined in the previous section, 
and used $L_{IR}/L_{UV}$ as an extinction indicator. 

The results for dust extinction using these three techniques is described in the next section.

 \subsubsection{Extinction Curve Difference in UV-wavelengths}

\begin{figure}
  \epsscale{1}\plotone{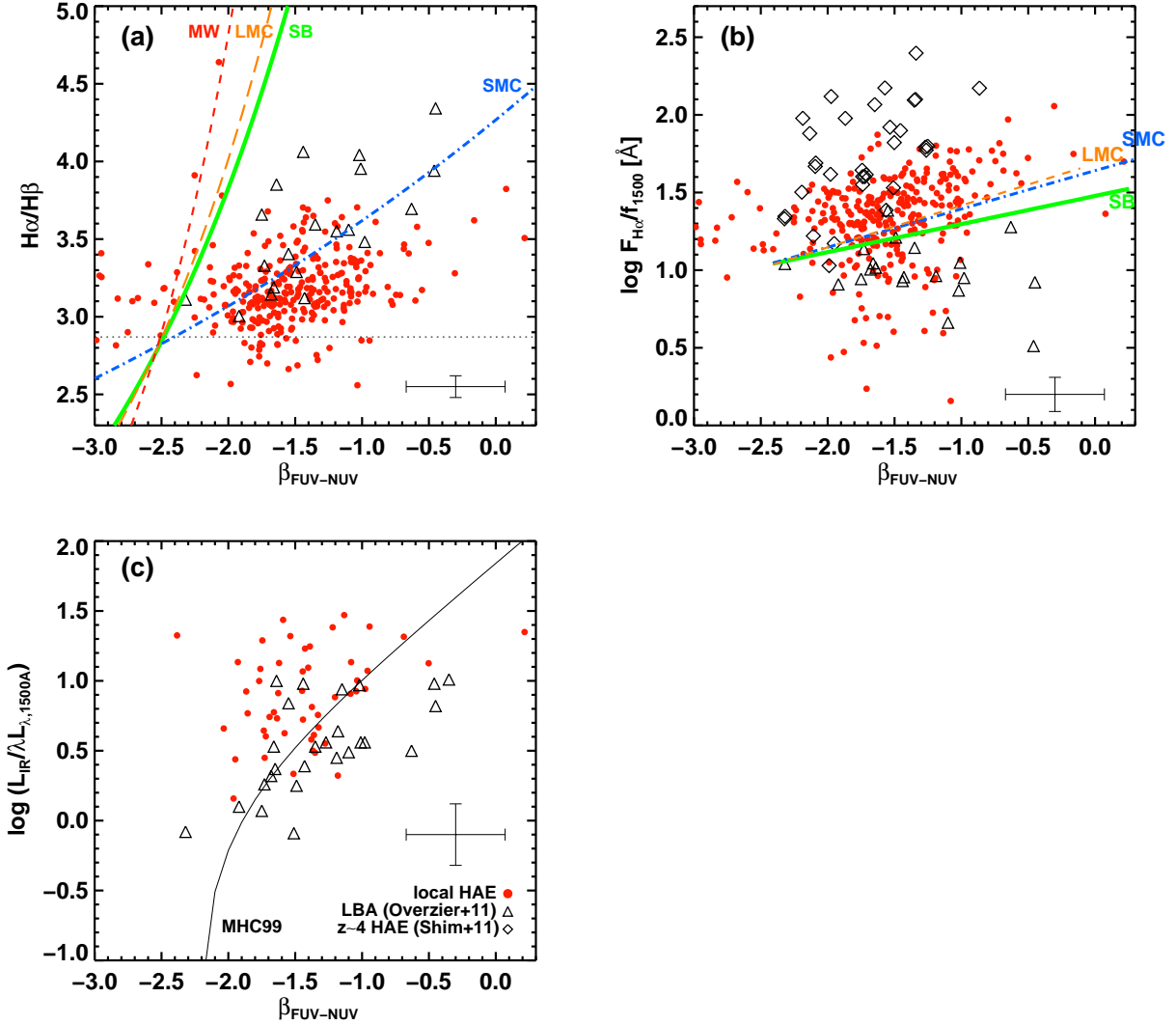}
  \caption{ \label{fig:ExtInd}
\footnotesize{(a) Balmer line ratio ($H\alpha/H\beta$) vs. UV spectral slope $\beta$
of local HAEs (filled circles)
and LBAs (triangles; Overzier et al. 2011).
UV slope $\beta$ is derived using the GALEX FUV-NUV colors (see text for details),
and $H\alpha/H\beta$ ratio are derived from the MPA-JHU catalog.
  Dotted horizontal line indicates $F_{H\alpha}/F_{H\beta}=2.87$,
  the unobscured Balmer line flux ratio for case-B recombination with
  $T=10^{4}$\,K (Osterbrock \& Ferland 2006).
Overplotted tracks indicate $H\alpha/H\beta$ vs. $\beta$ expected from 
different dust extinction laws (Allen 1976; Fitzpatrick 1986; Prevot et al. 1984; 
Calzetti et al. 2000). 
$H\alpha/H\beta$ ratio of local HAEs is difficult to explain using the SB extinction law: 
extinction laws with a steeper shape at UV wavelengths (e.g., SMC extinction law) 
is necessary. Some of the LBAs can be described using a SB extinction law with varying 
star formation history, but an SMC-like extinction law is required for
some of the LBAs. 
  (b) $H\alpha$ line-to-UV continuum flux ratio vs. UV spectral slope $\beta$
  of local HAEs (filled circles), LBAs (triangles),
  and $z\sim4$ HAEs (diamonds; Shim et al. 2011).
Here, $f_{1500}$ indicates $f_{\lambda}$ at $1500\,\mbox{\AA}$.
  Overplotted lines indicate the relation between $H\alpha$ line-to-UV continuum flux ratio
and $\beta$ assuming different extinction laws, which could be shifted along the y-axis
according to the star formation history assumed. 
However, H$\alpha$/UV ratio is affected by both extinction and stellar population age, 
thus it is not as effective as the $H\alpha/H\beta$ ratio as a probe of
different extinction laws among galaxies. 
Again in this plot, we see that there is almost no overlap between $z\sim4$ HAEs and LBAs. 
  (c) IR-to-UV luminosity ratio vs. UV spectral slope $\beta$
  of local HAEs and LBAs. Overplotted line shows
  the empirical IR-to-UV ratio vs. $\beta$ relationship
  for local starbursts (Meurer, Heckman, \& Calzetti 1999).
Considering that HAEs are young galaxies and LBAs are more likely to be evolved systems, 
the IR/UV ratio of LBAs are expected to be smaller than that of HAEs since 
the fraction of stellar radiation that is absorbed by dust and re-radiated in IR 
decreases as the galaxy evolves. The observed IR/UV ratio confirms this idea, 
showing that most HAEs have higher IR/UV ratio than LBAs or other local star-forming galaxies (line from MHC 1999).}
  }
 \end{figure}

Figure \ref{fig:ExtInd} illustrates the comparison between 
the three extinction indicators for local HAEs and LBAs. 
Figure \ref{fig:ExtInd}a shows that there exists a positive correlation 
between UV slope $\beta$ and $F_{H\alpha}/F_{H\beta}$, which is naturally
expected from the fact that both quantities are providing independent measures of obscuration.
However, the relationship between UV slope $\beta$ and $F_{H\alpha}/F_{H\beta}$
for local HAEs and LBAs are different, 
i.e., for the same $\beta$, the flux ratio $F_{H\alpha}/F_{H\beta}$ 
is higher for LBAs than the case of HAEs. 
One possibility is that the ISM temperature in LBAs and HAEs is different, with the latter having higher
gas temperatures which affects the recombination rate by a factor of 1.3.
This would be consistent with the higher ultraviolet surface density seen in the HAEs compared to LBAs.

The alternate interpretation is this is a result of dust extinction.  For a fixed $\beta_{FUV-NUV}$, LBAs show larger dust extinction
than HAEs, which implies that the UV dust extinction curve is steeper in HAEs than in LBAs. This
is more consistent with the extinction curve seen in the Small Magellanic Cloud (SMC)
which due to its low metallicity is thought to have an overabundance of large grains relative 
to small grains (Weingartner \& Draine 2001).

This possible difference in the shape of the dust extinction curve 
especially in the UV wavelength range has already been suggested in 
the study of $z\sim4$ HAEs (Shim et al. 2011). 
Figure \ref{fig:ExtInd}b shows the relationship between 
H$\alpha$ line-to-UV continuum ratio and UV slope $\beta$ for local HAEs,
in comparison with their high-redshift counterparts ($z\sim4$ HAEs)
and LBAs. For comparison, the figure also shows
three lines indicating the starburst extinction law (Calzetti 2000),
LMC extinction law (Fitzpatrick 1986), and SMC extinction law (Prevot 1984). 
The extinction laws that have a steeper curve 
at UV wavelengths (LMC, SMC) show higher H$\alpha$-to-UV ratio 
compared to less steep extinction law (SB). The $y$-intercepts in 
these model lines depend on the 
assumed star formation history and the stellar population age of the galaxy 
while only the slope of the lines depend on the extinction law (See Shim et al. 2011). 
Higher-redshift HAEs have higher H$\alpha$-to-UV ratio 
than local counterparts, implying that these are closer to 
continuously star-forming galaxies. Despite the $y$-intercept difference, 
HAEs at different redshifts show slopes 
in the H$\alpha$-UV ratio vs. $\beta$ relation which are consistent despite the
large scatter. On the other hand, 
LBAs appear to have a slope that is completely different in the
H$\alpha$-UV ratio vs. $\beta$ plane indicating that not only do they have distinct
extinction properties to HAEs but also different star-formation histories.

Finally in Figures \ref{fig:ExtInd}c, 
we present the IR-to-UV luminosity ratio for local HAEs. 
Since HAEs appear to be young galaxies 
while LBAs are relatively evolved galaxies, the
IR-to-UV ratio of LBAs are expected to be smaller than that of HAEs 
since the fraction of stellar radiation that is absorbed by dust
and re-radiated in the IR decreases as the galaxy evolves. 
The observed relation between the IR-to-UV ratio and UV spectral slope $\beta$
for local HAEs lie above the expected line for local star-forming galaxies 
(Meurer et al. 1999) or that of LBAs, confirming the idea that 
young HAEs have stronger intrinsic UV radiation fields and thereby emit more strongly in the IR. 
 Still, the uncertainties related to the derivation of $L_{IR}$
 and $\beta_{FUV-NUV}$ is large. The addition of far-infrared continuum flux
is needed to reduce the uncertainties in $L_{IR}$. 
To constrain the extinction curve shape in UV wavelengths in more detail, 
the accuracy in $\beta_{FUV-NUV}$ should also be improved with higher precision UV spectrophotometry which
disentangles the contribution of the Ly-$\alpha$ line to the broadband photometry.

\section{Summary}

 In this paper, we have studied the extremely rare, local analogs of $z\sim4$ star-forming galaxies from SDSS DR7.
It has recently been demonstrated in Shim et al. (2011)
that 70\,\% of $z\sim4$ galaxies have strong H$\alpha$ equivalent widths and are H$\alpha$ emitters (HAEs).
Therefore, unlike previous studies which selected local analogs based solely on their UV properties, we used
the criterion H$\alpha$ EW $>500\,\mbox{\AA}$ to select among the local galaxy population.
At $z<0.4$, the number fraction of such strong HAEs is only 0.04\,\%.
Local HAEs are less luminous by an order of magnitude in both H$\alpha$ and UV luminosities 
compared to $z\sim4$ HAEs.
However, the H$\alpha$-to-UV luminosity ratio as well as
the specific star-formation rates of local HAEs are very similar to those of high redshift HAEs. This
supports the argument that local HAEs are scaled down versions of
$z\sim4$ star-forming galaxies. 

In contrast,
previously studied UV-selected local analogs of high-redshift Lyman-break galaxies,
the Lyman-break analogs, 
are distinct in their physical properties from HAEs at any redshift.
The UV-selected analogs show a factor of 5 lower H$\alpha$ EW, 
higher metallicity and higher stellar mass than the HAEs.
At least 50\,\% of local HAEs show comparably 
high FUV surface brightness as that of Lyman break galaxies and Lyman break analogs.  
However, the FUV surface brightness does not appear to depend on H$\alpha$ EW. 
Unlike the Lyman break analogs,
the composite spectrum of local HAEs shows strong, but narrow Helium
lines that are observed in Wolf-Rayet galaxies, 
raising the possibility that the properties of the star forming environment
in local HAEs overlap with those of Wolf-Rayet galaxies. 

 As in the case of $z\sim4$ HAEs, 
 the strong H$\alpha$ emission in local HAEs can be attributed to a
 young stellar population with a large number of massive stars.
50\% of local HAEs show low metallicity
of $12+\mbox{log[O/H]}\sim8.0$,  close to $\sim0.1\,Z_{\odot}$ which 
is less than that of the UV-selected Lyman break analogs. This
is consistent
with the fact that HAEs are less massive by at least an order of magnitude than the Lyman break analogs. 
The low metallicities are also reflected in their low observed [NII]/H$\alpha$ ratio. 
The low [NII]/H$\alpha$ ratio, as well as high [OIII]/H$\beta$ ratio, 
indicate that local HAEs are not contaminated by AGN-dominated systems.
Moreover, the strong [OIII] emission 
that displaces local HAEs from the ridge line of BPT diagram 
can be explained by a high ionization parameter, 
which requires either higher electron densities 
and/or large escape fraction of hydrogen ionizing photons. 
If the large escape fraction is demonstrated through observational data, HAEs could be the clue
to understanding the reionization of the intergalactic medium at high redshift.

Most HAEs are unusually young with $D_n\,(4000) < 1$. 
This is one of the largest differences between the properties of
HAEs and Wolf-Rayet galaxies; the latter predominantly showing $D_n\,(4000) >1$ suggesting a more evolved stellar population. 
Due to their small $D_n\,(4000)$,
it is therefore difficult to constrain whether local HAEs follow a continuous star formation history
or burst-like star formation history. 
The extinction indicator H$\alpha$/H$\beta$ of local HAEs 
suggests that HAEs display an extinction curve steeper than normal star-forming galaxies and more similar to that of the Small Magellanic Cloud.
The origin of this difference in extinction curve is not clear with the most 
likely reason being the lower metallicities in these galaxies.

\begin{deluxetable}{l rrr r r r  rrc c c}  \tabletypesize{\scriptsize}
  \tablewidth{0pt}  
  \tablecaption{\label{tab:objlist} Local HAEs }
  \tablehead{
     \colhead{SDSS ID} & \colhead{R.A.} & \colhead{Decl.} & \colhead{redshift} & 
     \colhead{H$\alpha$ EW} & \colhead{$f(H\alpha)$} & {log\,$I_{FUV}$} &
     \colhead{log\,SFR}  & \colhead{log\,$M_*$} & \colhead{12$+$log[O/H]} & $D_n(4000)$
   }
  \startdata
    \tableline
587727179525783616 & 5.915092 & -9.813522 & 0.053035 & 667.5 & 4001.6 & 8.447 & 0.481 & 8.608 & 8.149 & 0.8980 &  \\
588015507661455435 & 6.104308 & -1.066375 & 0.039368 & 611.7 & 1085.2 & 8.784 & -0.610 & $\ldots$ & 8.007 & 0.9120 &  \\
588015507661979813 & 7.409042 & -1.204444 & 0.164415 & 752.8 & 865.7 & 9.116 & 0.675 & $\ldots$ & 8.155 & 0.8390 &  \\
588015508199506053 & 8.835831 & -0.746512 & 0.249825 & 503.4 & 773.8 & 8.819 & 0.920 & $\ldots$ & 8.255 & 0.9190 &  \\
587740588411519244 & 9.582756 & 25.219419 & 0.311599 & 657.4 & 597.0 & 8.841 & 1.073 & 9.278 & 8.412 & 0.9694 &  \\
587724199349387411 & 10.226349 & 15.569384 & 0.283232 & 636.7 & 478.4 & 8.939 & 0.873 & 9.240 & $\ldots$ & 0.8826 &  \\
587724234248552589 & 10.653869 & 16.034079 & 0.247397 & 844.6 & 995.4 & 11.839 & 1.668 & 9.346 & 8.809 & 0.8554 &  \\
587731186746196208 & 12.307815 & 0.400554 & 0.159282 & 896.6 & 580.6 & 9.116 & 0.256 & 8.942 & 7.673 & 0.7662 &  \\
587731514215366775 & 16.450899 & 1.080542 & 0.329321 & 574.2 & 273.3 & 7.890 & 0.851 & $\ldots$ & 8.751 & 1.2555 &  \\
587727180069142712 & 21.047823 & -9.002034 & 0.229820 & 913.6 & 489.1 & 11.317 & 0.942 & 9.071 & 7.876 & 0.9508 &  \\
587724232106508428 & 23.469007 & 13.702609 & 0.008670 & 1330.2 & 3342.0 & $\ldots$ & -1.394 & 6.598 & 7.672 & 0.0000 & WR \\
588015507669188698 & 23.856556 & -1.230119 & 0.177349 & 564.5 & 1235.2 & 8.980 & 1.068 & 9.289 & 8.489 & 1.0286 &  \\
588015508206190774 & 24.127474 & -0.632222 & 0.059468 & 505.1 & 972.5 & 8.868 & -0.407 & $\ldots$ & 8.127 & 0.9764 &  \\
587724232644820997 & 26.779322 & 13.941470 & 0.056623 & 684.3 & 3158.3 & 8.780 & 0.548 & 8.420 & 8.673 & 0.8630 &  \\
587731512610127975 & 28.668512 & -0.112153 & 0.018708 & 664.7 & 823.0 & 8.657 & -1.221 & $\ldots$ & 8.775 & 0.0000 &  \\
587731512610455795 & 29.539122 & -0.110334 & 0.012072 & 817.4 & 2152.9 & 8.856 & -1.412 & $\ldots$ & 7.933 & 0.0000 &  \\
587727779743596668 & 34.720455 & -9.205221 & 0.012708 & 686.3 & 2136.4 & 8.276 & -0.962 & 7.042 & 8.015 & 0.0000 &  \\
587727178464624784 & 35.156929 & -9.485357 & 0.113164 & 597.2 & 1955.8 & 9.439 & 0.525 & 9.272 & 9.270 & 0.8954 &  \\
587731514224279696 & 36.810349 & 1.093360 & 0.348540 & 1293.5 & 856.1 & 9.226 & 1.667 & $\ldots$ & 8.836 & 1.0014 &  \\
587727177929588906 & 39.490982 & -9.525620 & 0.280223 & 667.9 & 509.4 & 11.326 & 0.906 & 9.230 & $\ldots$ & 0.7763 &  \\
587727179003723785 & 40.217499 & -8.474285 & 0.082176 & 1795.5 & 4386.1 & 9.736 & 0.816 & 8.457 & 7.897 & 0.6837 &  \\
588015509824208968 & 40.995934 & 0.556302 & 0.059319 & 708.7 & 1001.1 & $\ldots$ & -0.432 & $\ldots$ & 8.038 & 0.8369 &  \\
587731514226442382 & 41.761272 & 1.258160 & 0.128720 & 844.0 & 1875.2 & 9.219 & 1.063 & 9.034 & 8.633 & 0.8386 &  \\
587727180078907412 & 43.444599 & -7.395548 & 0.004494 & 789.9 & 3784.6 & 9.951 & -2.010 & 6.062 & 7.953 & 0.0000 &  \\
587731511543726169 & 45.454273 & -0.882610 & 0.007318 & 758.9 & 1641.3 & 8.225 & -1.998 & $\ldots$ & 7.677 & 0.0000 &  \\
587727179006148758 & 45.839211 & -7.989777 & 0.164813 & 589.8 & 1817.3 & 9.327 & 0.838 & 9.121 & 7.817 & 0.8422 &  \\
587731514228474146 & 46.413918 & 1.189721 & 0.167591 & 859.9 & 453.8 & 8.129 & 0.281 & $\ldots$ & 8.045 & 0.9596 &  \\
588015509290483920 & 48.250191 & 0.103361 & 0.029164 & 577.7 & 1161.1 & 9.365 & -1.023 & $\ldots$ & 7.895 & 0.8494 &  \\
587744294975242477 & 48.357658 & 5.751317 & 0.195068 & 1259.5 & 993.4 & 8.972 & 1.493 & 9.035 & 8.151 & 0.9959 &  \\
587731512082170019 & 49.026310 & -0.438349 & 0.022894 & 563.0 & 1971.2 & 11.054 & -1.208 & 6.650 & 8.048 & 0.8465 &  \\
588015509290877040 & 49.099861 & 0.153407 & 0.202564 & 854.8 & 1144.3 & 9.785 & 0.882 & 9.236 & $\ldots$ & 0.7714 &  \\
587724241767825591 & 51.556793 & -6.586821 & 0.162033 & 638.1 & 1742.2 & 9.699 & 1.417 & 9.206 & 8.749 & 0.8595 &  \\
588015510365929716 & 52.101753 & 0.909903 & 0.276351 & 831.5 & 816.3 & 8.693 & 1.467 & $\ldots$ & 8.692 & 0.9247 &  \\
587731513157746887 & 53.330040 & 0.292035 & 0.193820 & 623.2 & 1025.7 & 9.241 & 0.937 & $\ldots$ & 8.364 & 0.9329 &  \\
587731514232275127 & 55.082912 & 1.058530 & 0.321582 & 530.0 & 314.4 & 11.201 & 0.889 & $\ldots$ & 8.710 & 1.1827 &  \\
587731511549427842 & 58.572277 & -0.913758 & 0.025706 & 602.7 & 2305.9 & 9.212 & -0.580 & $\ldots$ & 7.895 & 1.1308 &  \\
587727179550294225 & 62.406769 & -5.301616 & 0.074775 & 599.3 & 1696.9 & 9.217 & 0.212 & 8.472 & 8.041 & 0.8772 &  \\
758882136838308201 & 88.771126 & 83.197517 & 0.057179 & 514.5 & 1076.3 & $\ldots$ & -0.446 & 7.393 & 8.579 & 0.9396 &  \\
587725774528643229 & 113.236320 & 37.080456 & 0.139556 & 524.9 & 2160.2 & 9.060 & 1.242 & 9.764 & 8.687 & 0.9179 &  \\
587732054308094381 & 115.144371 & 24.693533 & 0.192887 & 564.0 & 1626.4 & $\ldots$ & 1.122 & 9.385 & 8.387 & 0.9499 &  \\
587732152555864324 & 116.991676 & 23.609142 & 0.155223 & 548.9 & 894.7 & $\ldots$ & 0.480 & 9.239 & 8.125 & 0.8992 &  \\
588016841241395565 & 117.026268 & 19.529713 & 0.062913 & 765.3 & 3074.9 & 9.269 & 0.469 & 8.640 & 8.620 & 0.9471 &  \\
587731887343141089 & 117.146057 & 31.510386 & 0.027381 & 642.6 & 4367.5 & 8.270 & -0.047 & 8.030 & 8.145 & 0.9204 &  \\
587731681185038501 & 118.126228 & 30.268797 & 0.104871 & 528.0 & 1131.6 & 9.115 & 0.471 & 9.145 & 8.011 & 0.9481 &  \\
588297865245360528 & 119.660004 & 25.432913 & 0.161057 & 533.9 & 2565.8 & $\ldots$ & 1.370 & 9.519 & 8.789 & 1.0312 &  \\
588007005769892050 & 120.446251 & 43.883945 & 0.084337 & 530.2 & 1589.3 & 9.124 & 0.337 & 8.968 & 8.573 & 0.9658 &  \\
587731885734625538 & 121.991661 & 34.244244 & 0.022444 & 1279.5 & 1851.5 & 8.252 & -0.946 & 6.787 & 7.905 & 0.7507 &  \\
587738947196944678 & 123.966682 & 21.939903 & 0.140995 & 852.4 & 1124.3 & 9.202 & 0.611 & 9.047 & 7.957 & 0.7882 &  \\
587732577221083492 & 125.895187 & 3.221021 & 0.009771 & 1355.1 & 16856.7 & 10.612 & -0.518 & 6.810 & 8.128 & 0.0000 &  \\
587735241709322314 & 125.979027 & 28.106043 & 0.047222 & 505.5 & 12206.2 & 10.336 & 1.474 & 8.608 & 8.833 & 0.9276 & WR \\
587725980151513279 & 126.377838 & 50.801239 & 0.096858 & 556.0 & 2543.5 & 9.488 & 0.623 & 8.864 & 8.477 & 0.8790 & WR \\
587741421636092149 & 126.418541 & 18.771444 & 0.037959 & 1283.9 & 2072.9 & 9.051 & -0.218 & 7.345 & 7.820 & 0.8439 &  \\
587731679041290347 & 126.481354 & 35.542213 & 0.002496 & 1441.0 & 6282.5 & 8.842 & -1.995 & 6.040 & 7.670 & 0.0000 &  \\
587739114701652062 & 127.664062 & 22.250998 & 0.016886 & 538.6 & 5476.6 & 8.798 & -0.304 & 7.951 & 8.425 & 0.9368 &  \\
588010137337200805 & 128.666870 & 48.094685 & 0.342574 & 890.6 & 929.5 & 9.218 & 2.055 & $\ldots$ & 8.782 & 0.8600 &  \\
587731680116867174 & 129.681839 & 38.897362 & 0.147461 & 620.0 & 4561.8 & 9.638 & 1.353 & 9.573 & 8.545 & 0.9363 & WR \\
587741489819025453 & 130.001556 & 18.091948 & 0.072188 & 824.2 & 2345.0 & 9.939 & 0.777 & 8.244 & 8.700 & 0.8078 &  \\
587742010046808228 & 130.142105 & 13.747596 & 0.226933 & 769.8 & 948.2 & 9.105 & 1.551 & 9.275 & 8.799 & 0.8952 &  \\
587745403070710009 & 130.652435 & 10.553878 & 0.010318 & 605.1 & 2504.7 & 8.219 & -1.191 & 7.010 & 7.847 & 0.0000 &  \\
588010358527951007 & 131.059311 & 2.439195 & 0.091093 & 560.0 & 4591.0 & $\ldots$ & 1.594 & 9.201 & 8.830 & 0.9673 & WR \\
587725471207260238 & 131.365067 & 53.148048 & 0.031071 & 631.5 & 5260.2 & 8.594 & 0.001 & 8.135 & 8.312 & 0.8630 & WR \\
587745539973382205 & 132.484436 & 10.719072 & 0.014143 & 506.8 & 7399.6 & 8.527 & -0.461 & 7.668 & 8.038 & 0.0000 &  \\
587737809037558040 & 132.815216 & 58.681950 & 0.091864 & 1384.0 & 3337.6 & 9.046 & 0.817 & 8.647 & 7.896 & 0.7205 &  \\
587745243620638850 & 133.090485 & 12.281044 & 0.075924 & 812.3 & 9556.6 & 9.418 & 0.983 & 8.986 & 8.418 & 0.8125 &  \\
587741532770074773 & 133.350372 & 19.506294 & 0.236503 & 564.4 & 629.5 & $\ldots$ & 0.819 & 9.173 & 8.176 & 0.8873 &  \\
587726031692103963 & 135.277344 & 0.905037 & 0.110702 & 507.4 & 1617.7 & 9.512 & 0.625 & 9.176 & 8.285 & 0.9508 & WR \\
587741421103611996 & 136.278595 & 22.642759 & 0.125548 & 500.3 & 1916.9 & 9.417 & 0.700 & 9.097 & 8.289 & 0.8951 &  \\
587731681193754681 & 136.367004 & 44.182873 & 0.065364 & 568.5 & 3355.1 & 9.069 & 0.546 & 8.702 & 8.526 & 0.8958 &  \\
588010359604052117 & 136.379486 & 3.591771 & 0.039088 & 591.5 & 2267.1 & 8.865 & -0.064 & 8.061 & 7.890 & 0.8622 & WR \\
587732048406249534 & 136.618698 & 44.981819 & 0.074672 & 566.1 & 2915.6 & 8.884 & 0.888 & 8.777 & 8.691 & 0.9416 &  \\
587734622171889671 & 138.037277 & 36.373997 & 0.165166 & 559.4 & 1500.0 & 9.169 & 1.119 & 9.231 & 8.828 & 0.8944 &  \\
587732049481039908 & 138.645645 & 47.035343 & 0.027269 & 569.9 & 8656.9 & 9.150 & 0.157 & 8.269 & 8.016 & 0.8824 &  \\
587741817314738195 & 139.170761 & 18.468864 & 0.021771 & 534.0 & 2973.1 & 9.026 & -0.595 & 7.438 & 8.040 & 0.8853 &  \\
587745243087372534 & 141.384872 & 14.053627 & 0.301211 & 533.1 & 659.8 & 9.364 & 1.192 & 9.164 & 8.267 & 0.8964 &  \\
588013384341913605 & 141.501694 & 44.460049 & 0.180667 & 578.0 & 1770.6 & 9.861 & 1.063 & 9.129 & 8.482 & 0.8847 &  \\
587742014876745993 & 141.869492 & 17.671837 & 0.288328 & 509.3 & 578.0 & 9.063 & 1.032 & 9.225 & 8.307 & 0.9100 &  \\
587734622173462551 & 142.026337 & 38.132477 & 0.060717 & 739.6 & 5650.9 & 8.841 & 0.723 & 8.723 & 8.541 & 0.8537 & WR \\
587725075530317872 & 142.326645 & 0.470347 & 0.093874 & 888.4 & 1866.7 & 9.511 & 0.444 & 8.560 & 8.001 & 0.7983 &  \\
587725817478840384 & 142.526810 & 60.448166 & 0.013659 & 587.4 & 8243.8 & 9.591 & -0.750 & 7.320 & 8.062 & 0.0000 & WR \\
587739114708402325 & 142.907227 & 29.333946 & 0.330326 & 641.7 & 720.0 & 9.944 & 1.885 & $\ldots$ & 8.848 & 1.0686 &  \\
588010136268505157 & 143.509933 & 55.239777 & 0.002565 & 894.8 & 5601.8 & 9.437 & -1.149 & 6.041 & 7.670 & 0.0000 &  \\
587734621637247125 & 143.911865 & 38.631794 & 0.137483 & 650.1 & 1823.5 & 8.848 & 1.082 & 9.260 & 8.656 & 0.8719 & WR \\
587735343188934969 & 144.097046 & 9.000285 & 0.223567 & 672.6 & 923.0 & 8.907 & 1.590 & 9.237 & 8.741 & 0.8825 &  \\
587741392646504460 & 144.213989 & 26.717690 & 0.294546 & 510.1 & 933.1 & 9.652 & 1.246 & 9.385 & 8.468 & 0.9041 &  \\
588017979952922757 & 145.726135 & 34.069969 & 0.022487 & 795.7 & 1233.7 & 8.691 & -1.102 & 6.833 & 7.836 & 0.8487 &  \\
587725073921278043 & 146.007797 & -0.642272 & 0.004810 & 1534.4 & 21890.4 & 8.757 & -1.049 & 6.690 & 7.867 & 0.0000 &  \\
587734623786238071 & 146.826477 & 41.637905 & 0.004658 & 870.0 & 3060.5 & 7.680 & -1.791 & 6.435 & 7.909 & 0.0000 &  \\
588848900972216400 & 147.597183 & 0.708133 & 0.097702 & 552.3 & 3004.3 & 10.388 & 1.087 & 8.785 & 8.787 & 0.8468 &  \\
587728930273493055 & 147.882355 & 52.993347 & 0.046263 & 951.8 & 5236.0 & $\ldots$ & 0.288 & 8.195 & 7.990 & 0.8205 &  \\
587727944033108132 & 148.112335 & 2.299958 & 0.119123 & 511.9 & 3007.5 & $\ldots$ & 1.274 & 9.307 & 8.768 & 0.9354 & WR \\
587732152033345685 & 149.076309 & 43.124393 & 0.275709 & 538.1 & 982.0 & 13.432 & 1.674 & 9.280 & 8.826 & 0.9439 &  \\
587742062133117119 & 151.835388 & 19.563786 & 0.031410 & 579.3 & 1430.3 & 8.933 & -0.657 & 7.454 & 7.949 & 0.8696 &  \\
587728309631975445 & 151.943817 & 2.874571 & 0.023492 & 580.8 & 4678.0 & 9.180 & -0.347 & 7.596 & 8.123 & 0.9071 & WR \\
587741817320505351 & 152.636688 & 22.011009 & 0.004226 & 1038.6 & 9742.0 & 9.281 & -1.563 & 6.525 & 7.928 & 0.0000 & WR \\
587735348561444896 & 152.677261 & 12.921337 & 0.061313 & 574.1 & 7544.8 & $\ldots$ & 0.733 & 8.679 & 8.548 & 0.8578 & WR \\
587745541055971461 & 152.747086 & 15.706535 & 0.055632 & 802.3 & 5601.7 & 9.488 & 0.479 & 8.393 & 7.958 & 0.8323 &  \\
587738410863493299 & 152.987869 & 13.139479 & 0.143776 & 1395.2 & 2139.5 & $\ldots$ & 0.808 & 9.241 & 7.988 & 0.6890 &  \\
587738409789751347 & 153.112595 & 12.343749 & 0.009565 & 807.2 & 8293.9 & $\ldots$ & -0.814 & 7.236 & 8.095 & 0.0000 & WR \\
587742061597032613 & 153.744339 & 19.538765 & 0.012629 & 615.4 & 1542.3 & $\ldots$ & -1.409 & 6.536 & 7.852 & 0.0000 &  \\
587739376706387983 & 153.859955 & 30.914406 & 0.091751 & 537.9 & 2986.5 & 9.991 & 1.065 & 9.017 & 8.812 & 0.9525 &  \\
587735661550698508 & 154.102158 & 37.912769 & 0.003879 & 597.0 & 10978.4 & 9.210 & -1.462 & 6.582 & 7.675 & 0.0000 & WR \\
587734861609566213 & 154.124542 & 7.568038 & 0.182909 & 575.6 & 1859.0 & 9.148 & 1.554 & 9.443 & 8.695 & 0.9242 &  \\
588017605211390138 & 154.513519 & 41.105858 & 0.237018 & 679.0 & 721.2 & 9.306 & 1.142 & 9.173 & 8.644 & 0.7943 &  \\
587733081878888643 & 154.731064 & 51.924393 & 0.129398 & 520.0 & 2320.1 & 8.893 & 1.003 & 9.342 & 8.154 & 0.9644 &  \\
587728881414897699 & 156.121887 & 5.414172 & 0.033191 & 525.6 & 6362.6 & 9.174 & 0.046 & 8.191 & 7.902 & 0.8815 &  \\
587732578845786234 & 157.912231 & 7.265705 & 0.252536 & 568.7 & 446.5 & $\ldots$ & 0.675 & 8.946 & 8.215 & 0.9299 &  \\
587741490367889543 & 158.112335 & 27.298689 & 0.192490 & 626.1 & 1302.3 & 9.891 & 1.233 & 9.286 & 8.697 & 0.9132 &  \\
587732134842531847 & 160.176544 & 49.206581 & 0.005011 & 1454.6 & 7045.5 & 7.843 & -1.505 & 6.329 & 8.030 & 0.0000 & WR \\
587742014884544559 & 160.289993 & 21.361889 & 0.003976 & 599.2 & 13890.7 & 9.532 & -1.273 & 6.766 & 7.962 & 0.0000 &  \\
588017979421622307 & 160.668793 & 37.917389 & 0.166793 & 612.9 & 2133.4 & $\ldots$ & 1.521 & 9.373 & 8.751 & 0.9133 &  \\
587728879269642285 & 161.240799 & 3.886987 & 0.012873 & 1464.5 & 10315.3 & 10.044 & -0.846 & 6.796 & 7.870 & 0.0000 &  \\
587734863223324735 & 161.335083 & 9.396972 & 0.054873 & 600.2 & 6184.5 & 8.380 & 0.897 & 8.873 & 8.544 & 0.9856 & WR \\
588848901515182119 & 161.478241 & 1.068288 & 0.026199 & 890.7 & 11053.6 & 9.170 & 0.119 & 8.056 & 8.547 & 0.7903 & WR \\
587738410330357781 & 161.724945 & 13.779382 & 0.010610 & 916.4 & 4409.3 & 8.882 & -1.153 & 6.806 & 8.042 & 0.0000 &  \\
587742863668412459 & 162.635483 & 15.635085 & 0.084427 & 1045.3 & 5962.5 & 9.416 & 0.894 & 8.649 & 8.062 & 0.8002 &  \\
587739096444698642 & 162.670166 & 34.496483 & 0.052196 & 560.4 & 3332.6 & 9.077 & 0.245 & 8.324 & 8.050 & 0.9103 &  \\
588017705070886933 & 163.321625 & 12.777442 & 0.021807 & 523.3 & 4132.4 & 8.103 & -0.219 & 8.070 & 8.072 & 0.8879 & WR \\
587742062138621980 & 164.967575 & 21.708441 & 0.115118 & 607.4 & 4306.8 & 8.260 & 1.183 & 9.716 & 8.590 & 0.9387 &  \\
588848900980015266 & 165.318329 & 0.804029 & 0.212857 & 1190.1 & 1025.2 & 10.368 & 1.059 & 9.144 & 8.000 & 0.8033 &  \\
587741600950845470 & 165.891815 & 25.404541 & 0.155750 & 600.4 & 3250.1 & 8.744 & 1.463 & 9.592 & 8.575 & 0.9247 &  \\
587741490907971706 & 166.243927 & 29.137699 & 0.002139 & 686.8 & 11346.6 & 10.537 & -2.520 & 6.109 & 7.994 & 0.0000 &  \\
587742015424233589 & 167.160431 & 22.636049 & 0.023815 & 630.1 & 4309.1 & 9.115 & -0.461 & 7.416 & 8.090 & 0.8827 &  \\
588017606290178168 & 169.363617 & 45.012020 & 0.184647 & 1267.6 & 859.8 & 7.988 & 0.821 & 9.276 & 8.144 & 0.7805 &  \\
587742775628005405 & 169.442932 & 17.740192 & 0.004916 & 538.2 & 6140.8 & 9.452 & -1.437 & 6.621 & 8.019 & 0.0000 &  \\
587726033317789755 & 169.686325 & 2.908351 & 0.020327 & 1284.1 & 4852.2 & 10.192 & -0.636 & 7.577 & 8.059 & 0.7053 &  \\
587739607548362788 & 170.779007 & 30.478922 & 0.005361 & 602.8 & 10808.8 & 10.285 & -1.248 & 6.154 & 8.689 & 0.0000 & WR \\
588009370688553009 & 172.130554 & 61.215439 & 0.084348 & 928.3 & 2451.9 & 9.144 & 0.849 & 8.680 & 8.649 & 0.8768 &  \\
587735696440885344 & 172.818207 & 57.066330 & 0.005561 & 609.6 & 1984.4 & 7.771 & -1.694 & 6.507 & 7.897 & 0.0000 &  \\
587732484357161015 & 173.690506 & 50.100925 & 0.026005 & 623.1 & 10756.5 & 9.864 & 0.013 & 8.047 & 8.421 & 0.8640 & WR \\
588017111292969009 & 174.099274 & 47.158077 & 0.010177 & 911.2 & 5198.8 & 9.470 & -1.221 & 6.675 & 7.934 & 0.0000 &  \\
587739408388980778 & 174.342255 & 35.407413 & 0.194313 & 562.8 & 2162.4 & 10.357 & 1.611 & 9.331 & 8.740 & 0.9274 &  \\
587742981247795242 & 175.087631 & 60.327374 & 0.146754 & 539.9 & 2470.4 & 9.891 & 0.873 & $\ldots$ & 8.653 & 0.8458 &  \\
587739647821807686 & 175.281204 & 32.427006 & 0.006015 & 601.2 & 9006.3 & 10.478 & -1.797 & 6.041 & 7.922 & 0.0000 &  \\
587739609697878126 & 175.951233 & 32.716095 & 0.073985 & 915.4 & 3054.6 & 9.168 & 0.147 & 8.550 & $\ldots$ & 0.8608 &  \\
588010879831113781 & 175.998093 & 5.365164 & 0.099142 & 596.1 & 1768.2 & 8.817 & 0.792 & 8.784 & 8.709 & 0.8537 &  \\
587748930312994906 & 176.705582 & 0.896104 & 0.056519 & 540.0 & 849.5 & 8.865 & -0.407 & 7.744 & 7.976 & 0.8131 &  \\
587742191512715306 & 177.113907 & 25.769936 & 0.045117 & 854.5 & 9305.0 & 9.024 & 0.550 & 8.507 & 8.061 & 0.8363 &  \\
587742775094345789 & 177.170303 & 17.942505 & 0.079110 & 919.6 & 4043.8 & 9.551 & 0.719 & 8.605 & 8.257 & 0.7842 &  \\
587739304214265947 & 177.268524 & 35.039696 & 0.021131 & 520.5 & 5706.4 & 8.944 & 0.064 & 7.707 & 8.670 & 1.0226 &  \\
587735348571996236 & 177.511383 & 15.023189 & 0.002448 & 834.3 & 22601.2 & $\ldots$ & -1.536 & 6.583 & 8.069 & 0.0000 & WR \\
587748928166101081 & 178.197968 & -0.668794 & 0.004619 & 590.0 & 3819.9 & 8.692 & -1.784 & 6.363 & 7.966 & 0.0000 &  \\
587739096987271268 & 178.512695 & 36.850090 & 0.268020 & 743.3 & 670.5 & 8.568 & 0.707 & 9.055 & $\ldots$ & 0.7893 &  \\
587731891114803229 & 178.868073 & 57.664436 & 0.017262 & 1023.5 & 15064.6 & 9.550 & -0.176 & 7.660 & 7.914 & 0.7802 & WR  \\
587735346962497616 & 180.139252 & 13.718885 & 0.066750 & 1110.9 & 6000.8 & 9.444 & 0.066 & 8.471 & $\ldots$ & 0.8387 & WR \\
587726033859248332 & 180.231827 & 3.401094 & 0.084712 & 777.2 & 2024.7 & $\ldots$ & 0.417 & 8.482 & 7.907 & 0.8343 &  \\
587726032248701055 & 180.342972 & 2.185653 & 0.003251 & 1014.1 & 3213.4 & 7.723 & -1.996 & 6.086 & 7.670 & 0.0000 &  \\
587741709958840329 & 180.457916 & 28.102964 & 0.055879 & 672.2 & 2718.8 & 9.224 & 0.169 & 8.241 & 8.039 & 0.8384 &  \\
587739408391471112 & 181.289520 & 35.830231 & 0.119502 & 746.0 & 2759.7 & 9.884 & 1.394 & 9.065 & 8.770 & 0.8169 &  \\
587741531715797096 & 181.343231 & 28.946814 & 0.107620 & 1303.6 & 1982.1 & 9.408 & 0.714 & 8.832 & 7.900 & 0.7614 &  \\
588011122502336742 & 181.772125 & 61.586613 & 0.262101 & 2270.8 & 448.2 & 12.980 & 0.723 & 9.048 & 8.181 & 0.8391 &  \\
587741491451002960 & 182.393707 & 30.890612 & 0.219277 & 574.2 & 1534.2 & 10.206 & 1.501 & 9.259 & 8.718 & 0.8622 &  \\
588017626154598555 & 182.534775 & 44.651821 & 0.022742 & 675.4 & 1456.2 & 8.300 & -0.899 & 7.243 & 8.028 & 0.9342 &  \\
588017567099388030 & 182.658813 & 13.023331 & 0.008039 & 873.0 & 2874.4 & 7.886 & -1.351 & 6.625 & 7.999 & 0.0000 &  \\
587742061609156755 & 183.351196 & 22.542812 & 0.162278 & 623.6 & 440.0 & 8.549 & 0.246 & 8.658 & 8.060 & 0.8466 &  \\
587733079202070556 & 183.510345 & 53.754841 & 0.003055 & 853.4 & 4177.1 & 8.413 & -1.939 & 6.025 & 7.707 & 0.0000 &  \\
587735696443506837 & 184.006226 & 57.577858 & 0.286866 & 1084.4 & 1180.1 & $\ldots$ & 1.242 & 9.038 & $\ldots$ & 0.8784 &  \\
587735349111947338 & 184.766602 & 15.435700 & 0.195579 & 1094.5 & 1714.8 & 9.846 & 1.129 & 9.266 & 7.893 & 0.7630 &  \\
587742012747022361 & 184.884247 & 21.556944 & 0.140986 & 645.4 & 3180.9 & 10.048 & 1.615 & 9.185 & 8.811 & 0.8737 &  \\
588298662500171878 & 185.610733 & 47.066746 & 0.186958 & 611.5 & 2524.3 & 9.532 & 1.762 & 9.568 & 8.796 & 0.9345 &  \\
587742954397630497 & 185.818893 & 4.836145 & 0.017845 & 520.5 & 5224.2 & 8.804 & -0.811 & $\ldots$ & 8.062 & 0.8640 &  \\
587739097526829082 & 186.152985 & 37.410152 & 0.040380 & 548.8 & 3170.1 & 9.265 & -0.095 & 7.864 & 7.907 & 0.8974 &  \\
588010878225088680 & 186.549576 & 4.260018 & 0.094233 & 984.5 & 2221.1 & 8.544 & 0.579 & $\ldots$ & 8.262 & 0.7379 &  \\
588017729763016856 & 187.033600 & 7.912061 & 0.068031 & 582.2 & 2076.4 & 8.842 & 0.062 & $\ldots$ & 8.250 & 0.9121 &  \\
588017702933758072 & 187.702515 & 12.045229 & 0.004185 & 577.7 & 5388.9 & 8.474 & -1.689 & 6.563 & 7.861 & 0.0000 &  \\
588848899379822768 & 189.216797 & -0.589419 & 0.008531 & 905.2 & 1317.8 & 9.443 & -1.916 & 6.129 & 8.045 & 0.0000 &  \\
588017728690257952 & 189.259445 & 6.925279 & 0.005381 & 1556.0 & 20058.4 & 10.698 & -0.819 & $\ldots$ & 8.707 & 0.0000 & WR \\
587735696981688461 & 189.505127 & 58.020718 & 0.083989 & 917.5 & 1786.0 & 8.612 & 0.443 & 8.464 & 8.403 & 0.8195 &  \\
588017109686943854 & 189.515656 & 46.305592 & 0.098789 & 740.7 & 2442.8 & 8.442 & 1.142 & 9.046 & 8.590 & 0.9575 &  \\
587732771595419747 & 189.528702 & 10.165563 & 0.003795 & 1106.1 & 6387.0 & 7.860 & -1.715 & 6.467 & 7.974 & 0.0000 &  \\
587725039022768267 & 190.497253 & -3.667338 & 0.009240 & 611.5 & 2370.1 & 8.569 & -1.425 & 6.590 & 7.896 & 0.0000 &  \\
588017977284427855 & 190.820267 & 38.769154 & 0.023091 & 531.9 & 2965.5 & 8.827 & -0.497 & 7.670 & 8.083 & 0.9677 &  \\
587726032253419628 & 191.097427 & 2.261253 & 0.239378 & 777.5 & 2103.5 & 9.553 & 1.631 & 9.570 & 8.688 & 0.8409 &  \\
588017979968847957 & 191.170593 & 40.708172 & 0.017520 & 967.9 & 2442.9 & 8.697 & -0.817 & 7.097 & 8.076 & 0.8386 &  \\
588017992294662200 & 191.287720 & 10.727820 & 0.165673 & 719.0 & 3626.7 & 9.292 & 1.616 & 9.564 & 8.770 & 0.8740 &  \\
587738570853974145 & 192.061798 & 15.974835 & 0.278174 & 622.2 & 768.9 & 8.756 & 1.311 & 9.323 & 8.644 & 0.9903 &  \\
588017570848768137 & 192.144318 & 12.567482 & 0.263408 & 669.4 & 881.5 & 10.583 & 1.002 & 9.127 & 8.255 & 0.9013 &  \\
587732483288989834 & 193.416092 & 49.516125 & 0.231133 & 513.5 & 426.3 & 9.095 & 0.589 & 8.964 & 8.133 & 0.9404 &  \\
587724649799942319 & 193.858643 & -2.226135 & 0.051885 & 1278.5 & 1582.4 & 9.196 & -0.327 & 7.667 & $\ldots$ & 0.7685 &  \\
587726014538383550 & 194.380783 & 1.933281 & 0.252218 & 577.6 & 560.5 & 11.321 & 1.238 & 9.232 & 8.598 & 1.0160 &  \\
587738950417448968 & 195.763855 & 35.857948 & 0.060287 & 571.7 & 4688.5 & $\ldots$ & 0.573 & 8.741 & 8.526 & 0.9056 &  \\
587739098067107878 & 195.976852 & 37.233856 & 0.035953 & 518.6 & 5864.9 & 9.103 & 0.186 & 8.335 & 8.562 & 0.8792 & WR \\
587739304758083615 & 196.600800 & 35.228622 & 0.016196 & 650.3 & 5764.7 & $\ldots$ & -0.438 & 7.516 & 8.017 & 0.8456 &  \\
587742062151467120 & 196.734818 & 22.694004 & 0.274104 & 521.6 & 451.1 & 8.173 & 0.887 & 9.228 & 8.326 & 0.9161 &  \\
587733080816091172 & 196.869583 & 54.447128 & 0.032524 & 544.9 & 8271.5 & 10.657 & 0.005 & 7.781 & 8.568 & 0.9184 & WR \\
588848899383623800 & 197.812668 & -0.466085 & 0.230964 & 575.0 & 801.3 & 8.759 & 1.420 & 9.357 & 8.693 & 0.9482 &  \\
587741724973924354 & 199.480362 & 23.261173 & 0.024026 & 582.0 & 2775.0 & 8.332 & -0.357 & 7.912 & 8.086 & 0.9145 &  \\
587724648191885354 & 199.604355 & -3.418798 & 0.128790 & 617.3 & 2723.1 & 9.483 & 1.544 & 9.445 & 8.930 & 0.9964 &  \\
587733080280203402 & 200.460190 & 53.691071 & 0.032849 & 620.9 & 1215.2 & $\ldots$ & -0.483 & 7.600 & 8.103 & 0.8455 &  \\
587729775021981805 & 200.947754 & -1.547763 & 0.022464 & 1460.7 & 4136.0 & 9.764 & -0.727 & 7.042 & 7.870 & 0.7238 &  \\
587742062153236558 & 201.152542 & 22.183628 & 0.075391 & 747.1 & 2376.6 & 9.070 & 0.446 & 8.421 & 8.485 & 0.8791 &  \\
587729385546186821 & 201.229340 & 57.752960 & 0.116320 & 528.5 & 1764.5 & 8.590 & 0.928 & 9.268 & 8.754 & 0.9172 &  \\
587735429081530430 & 201.332870 & 48.040596 & 0.016369 & 669.9 & 3471.1 & 8.694 & -0.717 & 7.423 & 8.024 & 0.8676 &  \\
587739406251327528 & 201.455933 & 33.065105 & 0.014620 & 534.3 & 9028.6 & 9.043 & -0.458 & 7.720 & 8.359 & 0.9279 &  \\
587726031184330774 & 201.727585 & 1.229622 & 0.179626 & 679.3 & 2060.7 & 9.441 & 1.060 & 9.296 & 8.075 & 0.9287 &  \\
588017947210154051 & 201.847031 & 40.367821 & 0.010498 & 688.7 & 1261.1 & 8.706 & -1.700 & 6.273 & 7.802 & 0.0000 &  \\
588017605226594329 & 202.183563 & 43.930698 & 0.027969 & 559.8 & 14970.0 & 9.654 & 0.333 & 8.336 & 8.610 & 0.8825 & WR \\
587742773494218867 & 202.220657 & 15.992878 & 0.022747 & 866.1 & 3471.1 & 9.281 & -0.680 & 7.179 & 7.845 & 0.8386 &  \\
587742903405379689 & 202.318985 & 17.005835 & 0.094252 & 738.3 & 4768.0 & 10.383 & 1.328 & 8.970 & 8.811 & 0.8630 &  \\
587742774031220880 & 202.470200 & 16.342064 & 0.180808 & 778.9 & 1996.8 & 9.298 & 1.118 & 9.398 & 8.361 & 0.8456 &  \\
588017948284092523 & 202.536865 & 41.123695 & 0.027297 & 539.9 & 971.7 & 8.266 & -0.675 & 7.280 & 8.005 & 0.9497 &  \\
588017949357834271 & 202.862137 & 41.863415 & 0.011701 & 744.3 & 5782.7 & 8.741 & -0.885 & 7.159 & 7.854 & 0.0000 &  \\
587733411516842159 & 203.692856 & 53.824303 & 0.167942 & 502.2 & 888.1 & 8.726 & 1.007 & 9.018 & 8.781 & 0.9338 &  \\
587742550137241839 & 203.757736 & 18.527227 & 0.314059 & 679.4 & 465.2 & 9.626 & 0.940 & 9.044 & $\ldots$ & 0.9516 &  \\
588017726556143757 & 203.906677 & 8.030304 & 0.123465 & 847.5 & 1983.8 & 9.466 & 0.755 & 8.982 & 7.996 & 0.8346 &  \\
587738953104556058 & 203.950333 & 37.029358 & 0.056500 & 586.5 & 2891.1 & $\ldots$ & 0.482 & 8.441 & 8.553 & 0.9542 &  \\
587738570859413642 & 204.867935 & 15.278371 & 0.192017 & 932.6 & 1808.1 & 9.872 & 1.609 & 9.298 & 8.676 & 0.7877 &  \\
587739608635146287 & 205.485336 & 30.519339 & 0.003196 & 533.2 & 10491.2 & 8.374 & -1.504 & 6.979 & 8.107 & 0.0000 & WR \\
587735666377883763 & 206.114014 & 56.024929 & 0.070633 & 687.8 & 6400.2 & 8.839 & 1.033 & 9.146 & 8.549 & 0.9120 & WR \\
587739609709346918 & 206.848175 & 31.215101 & 0.119167 & 569.9 & 1492.4 & 8.979 & 0.552 & 9.374 & 8.269 & 0.9133 &  \\
587739811557867563 & 207.029114 & 26.405453 & 0.057039 & 555.6 & 4357.2 & 8.697 & 0.576 & 8.807 & 8.147 & 0.9111 &  \\
587742062693187673 & 208.642700 & 21.831629 & 0.110722 & 615.3 & 1465.5 & $\ldots$ & 0.663 & 9.765 & 8.289 & 0.9298 &  \\
588848901536481289 & 210.078918 & 1.081637 & 0.121090 & 865.0 & 2392.8 & $\ldots$ & 0.921 & 8.991 & 8.564 & 0.8628 &  \\
587735695913386093 & 211.119293 & 54.397999 & 0.000970 & 1126.5 & 17817.4 & $\ldots$ & -1.935 & 6.036 & 8.742 & 0.0000 &  \\
587726102556377282 & 211.855453 & 5.477178 & 0.084972 & 579.5 & 1425.3 & 9.111 & 0.166 & 8.444 & 8.002 & 0.8908 &  \\
587735696987389965 & 212.486496 & 54.946915 & 0.077313 & 721.9 & 6030.7 & 9.581 & 1.364 & 8.988 & 8.787 & 0.8831 & WR \\
588298661971820653 & 212.746735 & 43.046371 & 0.065605 & 613.0 & 4279.9 & 10.072 & 1.277 & 8.449 & 8.752 & 0.9095 &  \\
587742551751721066 & 213.140366 & 18.500360 & 0.007383 & 568.7 & 3461.5 & $\ldots$ & -1.410 & 6.882 & 7.994 & 0.0000 &  \\
587735696987717870 & 213.630035 & 54.515598 & 0.227058 & 508.9 & 393.8 & 9.521 & 0.515 & 9.033 & 8.201 & 0.8706 &  \\
587739380986216524 & 214.643311 & 31.211664 & 0.163290 & 531.8 & 433.5 & 9.977 & 0.310 & 8.572 & 7.891 & 0.9107 &  \\
587739843771695144 & 214.713028 & 21.044373 & 0.008547 & 1156.2 & 6710.1 & $\ldots$ & -1.161 & 6.632 & 7.674 & 0.0000 &  \\
587722982292521071 & 215.559616 & -0.655451 & 0.105870 & 572.2 & 1330.7 & 9.785 & 0.268 & 8.729 & 8.016 & 0.8116 &  \\
587739826591563799 & 215.928650 & 22.957996 & 0.032848 & 743.2 & 4621.7 & 9.421 & -0.189 & 7.650 & 7.805 & 0.8238 &  \\
588017114517536797 & 216.023865 & 42.279533 & 0.184816 & 1083.9 & 2550.9 & 10.255 & 1.675 & 9.284 & 8.787 & 0.7707 &  \\
587728920058200166 & 216.579788 & 62.463882 & 0.111854 & 584.1 & 1431.6 & 9.294 & 0.403 & 9.078 & 8.574 & 0.8677 & WR \\
588017948825813029 & 216.617355 & 38.382965 & 0.022302 & 535.4 & 2074.6 & 9.449 & -0.805 & 7.189 & 7.939 & 0.8890 &  \\
588017979977826415 & 217.022964 & 36.452888 & 0.086236 & 530.5 & 3105.9 & 8.794 & 0.574 & 8.841 & 8.022 & 0.8709 &  \\
587736525374357550 & 217.445862 & 6.726379 & 0.173502 & 850.2 & 4236.5 & 10.481 & 1.985 & 9.445 & 8.810 & 0.8173 &  \\
587726031728345220 & 218.042328 & 1.431060 & 0.136353 & 619.1 & 1768.1 & 9.372 & 1.142 & 9.005 & 8.808 & 0.9050 &  \\
587742577530372265 & 219.488220 & 17.322380 & 0.201980 & 1069.5 & 950.0 & 9.659 & 0.766 & 9.215 & $\ldots$ & 0.7740 &  \\
587726014549328120 & 219.517548 & 1.559333 & 0.312374 & 534.2 & 747.9 & 10.290 & 1.790 & 9.366 & 8.841 & 0.9448 &  \\
587736545772044617 & 220.114777 & 5.531906 & 0.005195 & 976.7 & 2689.7 & 8.539 & -1.817 & 6.170 & 7.789 & 0.0000 &  \\
587729777446945029 & 220.630737 & -2.164453 & 0.293644 & 899.7 & 830.3 & 10.028 & 1.118 & 9.112 & 7.999 & 0.7170 &  \\
587726101486764162 & 221.172363 & 4.161592 & 0.038811 & 841.9 & 1110.9 & 7.717 & -0.702 & 7.359 & 7.679 & 0.8334 &  \\
588011218602623163 & 221.949203 & 57.083385 & 0.285271 & 641.3 & 499.2 & 8.609 & 1.074 & 8.997 & 8.865 & 0.8904 &  \\
587729778521342037 & 222.022400 & -1.182697 & 0.027412 & 792.0 & 14736.6 & 9.634 & 0.475 & 8.090 & 7.951 & 0.8693 &  \\
587733398637314061 & 222.735794 & 48.623966 & 0.091899 & 536.6 & 4395.6 & 9.596 & 0.973 & 9.170 & 8.539 & 0.8913 &  \\
588018055652769997 & 223.648239 & 45.482330 & 0.268552 & 505.7 & 786.5 & 9.707 & 1.692 & 9.269 & 8.927 & 1.0017 &  \\
588023721783328799 & 223.690521 & 20.527960 & 0.015779 & 508.9 & 2959.8 & $\ldots$ & -0.984 & 7.179 & 7.992 & 0.0000 &  \\
588017626705166406 & 223.775284 & 38.137966 & 0.027702 & 703.9 & 5563.8 & 9.567 & 0.060 & 7.601 & 8.627 & 0.8179 &  \\
587739828742389914 & 224.396408 & 22.533831 & 0.148653 & 923.9 & 2325.5 & 9.286 & 1.009 & 9.241 & 8.296 & 0.7807 &  \\
587736976343171214 & 224.484177 & 30.862373 & 0.076537 & 539.6 & 1927.9 & 8.947 & 0.704 & 8.536 & 8.692 & 0.9438 &  \\
588011103712706632 & 226.616592 & 56.450748 & 0.278579 & 502.9 & 742.5 & 8.949 & 1.691 & 9.445 & 8.899 & 1.0494 &  \\
587729227147968579 & 226.869537 & 59.986984 & 0.182027 & 542.6 & 1727.1 & 9.325 & 1.533 & 9.436 & 8.757 & 0.9745 & WR \\
587733397565341781 & 227.287689 & 45.719120 & 0.048121 & 558.7 & 4671.8 & 8.916 & 0.349 & 8.359 & 8.569 & 0.9481 & WR \\
588017627780153375 & 227.392395 & 37.529476 & 0.032549 & 1348.0 & 6715.9 & 8.947 & 0.329 & 7.784 & 7.921 & 0.7920 &  \\
587733399712628848 & 228.053574 & 47.275188 & 0.053158 & 564.5 & 9973.9 & 8.715 & 1.012 & 9.119 & 8.611 & 0.9387 & WR \\
587739629560332485 & 228.229614 & 24.219585 & 0.073510 & 539.8 & 1591.2 & 9.309 & 0.114 & 8.389 & 8.040 & 0.8804 &  \\
587733412061577276 & 229.483307 & 46.418430 & 0.019327 & 594.6 & 2838.0 & 8.866 & -0.767 & 7.241 & 8.003 & 0.8990 &  \\
588011218605179021 & 230.190155 & 53.635342 & 0.207637 & 725.7 & 629.5 & 9.168 & 0.683 & 9.234 & 8.407 & 0.9447 &  \\
587739845925732735 & 230.661835 & 18.914272 & 0.182822 & 875.9 & 509.3 & $\ldots$ & 0.619 & 9.171 & 7.998 & 0.8893 &  \\
588017977836240925 & 230.884140 & 29.520021 & 0.068164 & 780.7 & 3333.1 & 9.164 & 0.344 & 8.405 & 7.986 & 0.8220 &  \\
587736976882663541 & 230.962494 & 28.616953 & 0.085434 & 620.5 & 1934.2 & 9.911 & 0.473 & 8.582 & 8.614 & 0.8938 &  \\
587732484375314456 & 232.071594 & 39.947338 & 0.064470 & 633.4 & 2909.7 & 8.869 & 0.449 & 8.753 & 8.552 & 0.8312 & WR \\
588848900472373334 & 232.128006 & 0.294496 & 0.113605 & 505.2 & 2186.0 & 9.149 & 0.448 & 9.273 & 7.670 & 0.9574 &  \\
587736618784587856 & 232.671951 & 31.018482 & 0.087004 & 516.8 & 4146.2 & $\ldots$ & 0.934 & 9.243 & 8.300 & 0.9380 &  \\
587742577536205016 & 232.982117 & 14.321038 & 0.031074 & 728.8 & 1892.1 & 8.156 & -0.424 & 7.549 & 7.824 & 0.8503 &  \\
587742061093322861 & 233.639023 & 14.913211 & 0.073269 & 600.4 & 4016.0 & 9.533 & 0.597 & 8.758 & 8.134 & 0.9021 &  \\
587729229297090692 & 234.405319 & 58.794586 & 0.214291 & 722.0 & 668.8 & 8.643 & 0.830 & 9.097 & 8.485 & 0.9031 &  \\
587735666387058818 & 235.333435 & 45.605324 & 0.202897 & 800.5 & 1571.6 & 10.129 & 1.627 & 9.269 & 8.757 & 0.9311 &  \\
587736915148996802 & 236.431473 & 8.967041 & 0.037724 & 1086.2 & 13056.6 & 10.073 & 0.492 & 8.041 & 7.841 & 0.7845 &  \\
587726101493579972 & 236.727264 & 3.150609 & 0.208937 & 613.6 & 2060.5 & 9.310 & 1.313 & 9.545 & 8.545 & 0.9161 &  \\
587736542027710819 & 236.755005 & 5.807413 & 0.230501 & 635.0 & 719.2 & 9.296 & 1.350 & 9.206 & 8.779 & 0.8723 &  \\
587726102030451047 & 236.787933 & 3.603900 & 0.231209 & 777.0 & 710.0 & 8.450 & 1.242 & 9.259 & 8.659 & 0.9510 &  \\
587739720845230356 & 237.615784 & 19.349661 & 0.212293 & 551.9 & 1531.4 & 9.688 & 1.874 & 9.422 & 8.868 & 0.8484 &  \\
587739652107600089 & 238.041687 & 21.053413 & 0.233148 & 555.8 & 628.5 & 9.161 & 0.923 & 9.193 & 8.264 & 0.8091 &  \\
587742061095551365 & 238.709274 & 13.435967 & 0.148752 & 901.5 & 1005.4 & 8.679 & 0.800 & 9.127 & 8.344 & 0.8277 &  \\
587736975276048497 & 239.384613 & 23.347342 & 0.032834 & 732.1 & 13564.8 & 9.772 & 0.531 & 8.049 & 8.651 & 0.9593 & WR \\
587733399717413034 & 239.935684 & 40.556999 & 0.226811 & 500.1 & 682.6 & 9.522 & 0.828 & 9.139 & 8.269 & 0.8699 &  \\
587736915687964980 & 241.152771 & 8.333086 & 0.312282 & 662.1 & 552.5 & 9.240 & 0.942 & 9.183 & 8.041 & 0.9067 &  \\
587729409147666501 & 241.417877 & 50.754395 & 0.012880 & 600.7 & 2121.6 & 9.534 & -1.437 & 6.275 & 8.084 & 0.0000 &  \\
587739844320166055 & 241.614731 & 13.929967 & 0.107036 & 780.1 & 3672.8 & 10.099 & 1.296 & 8.866 & 8.711 & 0.8247 &  \\
588018253759054025 & 242.043182 & 35.469250 & 0.032705 & 1296.1 & 1960.0 & 8.993 & -0.382 & 7.309 & 7.870 & 0.7605 &  \\
587739385278890314 & 243.373352 & 20.090286 & 0.051053 & 546.5 & 726.2 & 8.791 & -0.481 & 7.610 & 7.978 & 0.9221 &  \\
587736976351822098 & 244.027786 & 21.638210 & 0.288249 & 529.5 & 629.6 & 9.044 & 0.623 & 9.099 & $\ldots$ & 0.8935 &  \\
587742628534354235 & 244.070374 & 9.354526 & 0.048400 & 562.3 & 2155.5 & 8.338 & 0.077 & 8.189 & 8.110 & 0.9804 &  \\
587739706886062320 & 245.469055 & 15.315551 & 0.034341 & 672.1 & 7771.3 & 8.976 & 0.711 & 8.377 & 8.711 & 1.1473 &  \\
587745968930029927 & 245.681107 & 16.605579 & 0.149523 & 542.6 & 1911.6 & 10.843 & 1.595 & $\ldots$ & 8.794 & 1.0122 &  \\
587722982842761249 & 246.042145 & -0.367383 & 0.031327 & 649.5 & 14186.6 & 9.371 & 0.483 & 8.298 & 8.257 & 0.8995 &  \\
587739810501361893 & 246.963211 & 13.587149 & 0.016464 & 540.5 & 5675.7 & 9.042 & -0.347 & 7.776 & 8.577 & 0.9545 & WR \\
587739720849555951 & 247.051987 & 15.199133 & 0.316043 & 576.9 & 416.7 & 8.978 & 0.900 & 9.181 & $\ldots$ & 0.9590 &  \\
588018253224935626 & 247.114746 & 30.914915 & 0.114333 & 567.3 & 1605.6 & 9.261 & 0.074 & 9.064 & 7.670 & 0.9250 &  \\
587729408620495044 & 258.152618 & 32.275951 & 0.011950 & 797.3 & 3946.4 & $\ldots$ & -1.065 & 7.050 & 7.868 & 0.0000 &  \\
587725503949177426 & 260.040924 & 54.359188 & 0.293770 & 790.4 & 321.1 & 8.847 & 0.622 & 8.875 & $\ldots$ & 0.8666 &  \\
587725576962244831 & 261.776398 & 59.817295 & 0.347100 & 502.7 & 522.0 & $\ldots$ & 1.663 & $\ldots$ & $\ldots$ & 0.9932 &  \\
587725576426618961 & 262.277313 & 56.888676 & 0.015791 & 503.9 & 5343.1 & 8.911 & -0.310 & 7.715 & 8.969 & 1.0180 & WR \\
587725577500360837 & 263.755188 & 57.052376 & 0.047232 & 1347.3 & 17026.0 & 9.705 & 1.155 & 8.596 & 8.770 & 0.7995 & WR \\
587727214417871520 & 315.309967 & -5.919530 & 0.196179 & 561.8 & 1578.1 & 9.207 & 1.010 & 9.485 & 8.368 & 0.9205 &  \\
587731174918062687 & 316.967316 & 1.214631 & 0.175984 & 689.2 & 660.8 & 8.644 & 0.467 & 9.036 & 8.013 & 0.8687 &  \\
587726877265559558 & 318.862823 & -7.997601 & 0.028449 & 629.7 & 4729.1 & 9.666 & -0.199 & 7.616 & 8.082 & 0.9073 &  \\
587730774410068219 & 323.657532 & 11.419499 & 0.022039 & 578.3 & 5204.3 & 10.794 & -0.998 & 6.831 & 8.429 & 0.9261 & WR \\
587731187802767922 & 333.112244 & 1.143138 & 0.210114 & 561.5 & 465.3 & 8.104 & 0.617 & $\ldots$ & $\ldots$ & 0.9567 &  \\
587734304877707550 & 333.179413 & 0.113483 & 0.177042 & 615.3 & 696.9 & 9.671 & 0.612 & $\ldots$ & 8.572 & 0.9595 &  \\
587734304878035106 & 333.846069 & 0.046320 & 0.077447 & 769.5 & 2854.2 & 9.130 & 1.015 & $\ldots$ & 8.711 & 0.8438 &  \\
587731186193530964 & 336.292206 & -0.198011 & 0.066681 & 724.3 & 4011.3 & 9.590 & 0.437 & 8.401 & 7.979 & 0.8254 &  \\
587727222478667784 & 339.629669 & 14.008272 & 0.020607 & 955.9 & 3569.8 & 9.742 & -0.769 & 6.721 & 7.713 & 0.8022 &  \\
587731187271336095 & 345.541687 & 0.827455 & 0.033118 & 853.6 & 2106.8 & 8.976 & -0.558 & $\ldots$ & 7.727 & 0.8257 &  \\
587731187810042042 & 349.763550 & 1.148190 & 0.030129 & 575.0 & 1857.7 & 8.859 & -0.469 & $\ldots$ & 7.953 & 0.9916 &  \\
587731186201460957 & 354.466370 & -0.166820 & 0.071730 & 727.8 & 1832.8 & 9.203 & 0.296 & $\ldots$ & 8.019 & 0.8440 &  \\
588015510341943481 & 357.280731 & 0.932718 & 0.185753 & 511.3 & 441.1 & 9.487 & 0.215 & $\ldots$ & 8.152 & 0.9902 &  \\
   \enddata
   \tablecomments{Col.(1): SDSS object ID.
Col.(2): Right Ascension (J2000). Col.(3): Declination (J2000).
Col.(4): Redshift. Col.(5) H$\alpha$ equivalent width in units of $\mbox{\AA}$, 
from MPA-JHU value added catalog for SDSS DR7. 
Col.(6): H$\alpha$ line flux in units of $10^{-17}$ erg s$^{-1}$ cm$^{-2}$, 
from MPA-JHU value added catalog for SDSS DR7.
Col.(7): FUV luminosity surface density in units of $L_{\odot}$ kpc$^{-2}$.
FUV luminosity is derived using GALEX FUV magnitude, and is divided with 
$2\pi r_{hl}^2$, while $r_{hl}$ is half-light radius in SDSS $u$-band.
Col.(8): log of the star formation rate in units of $M_{\odot}$ yr$^{-1}$,
from MPA-JHU value added catalog for SDSS DR7.
Col.(9): log stellar mass in units of $M_{\odot}$ from MPA-JHU value added catalog for SDSS DR7.
Col.(10): 12$+$log[O/H] from MPA-JHU value added catalog for SDSS DR7.
Col.(11): $D_n(4000)$ from MPA-JHU value added catalog for SDSS DR7.
Col.(12): Comment for each object. We marked objects that are matched with Wolf-Rayet
galaxies (Brinchmann et al. 2008) as `WR'. 
    }
\end{deluxetable}

\end{document}